\documentclass{pasa}%

\usepackage{graphicx}
\usepackage{longtable}
\usepackage{ltablex}
\usepackage{comment}
\usepackage{epstopdf}
\usepackage{threeparttable}
\newcommand{\arcsec}{^{\prime \prime}}
\def\@mcsc{ptmrc}

\title[SkyMapper Southern Survey DR1]{SkyMapper Southern Survey: First Data Release (DR1)}
\author[Wolf et al.]{Christian Wolf$^{1,2}$, Christopher A. Onken$^{1,2}$, Lance C. Luvaul$^1$, Brian P. Schmidt$^{1,2}$, Michael S. Bessell$^1$, Seo-Won Chang$^{1,2}$, Gary S. Da Costa$^1$, Dougal Mackey$^1$, Tony Martin-Jones$^1$, Simon J. Murphy$^{1,3}$, Tim Preston$^{1,4}$, Richard A. Scalzo$^{1,2,5}$, Li Shao$^{1,6}$, Jon Smillie$^7$, Patrick Tisserand$^{1,8}$, Marc C. White$^1$ and Fang Yuan$^{1,2,9}$
\affil{$^1$Research School of Astronomy and Astrophysics, Australian National University, Canberra, ACT 2611, Australia}
\affil{$^2$ARC Centre of Excellence for All-sky Astrophysics (CAASTRO)}
\affil{$^3$Present address: School of Physical, Environmental and Mathematical Sciences, University of New South Wales Canberra, ACT 2600, Australia}
\affil{$^4$Present address: Carolus Software Ltd., 244 5th Avenue, New York, NY, 10001, USA}
\affil{$^5$Present address: Centre for Translational Data Science, University of Sydney, NSW 2006, Australia}
\affil{$^6$Kavli Institute for Astronomy and Astrophysics, Peking University, 5 Yiheyuan Road, Haidian District, Beijing 100871, P.~R.~China}
\affil{$^7$National Computational Infrastructure, Australian National University, Canberra ACT 2601, Australia}
\affil{$^8$Present address: Sorbonne Universit\'{e}s, UPMC Univ Paris 6 et CNRS, Institut d`Astrophysique de Paris, 98 bis bd Arago, F-75014 Paris, France}
\affil{$^9$Present address: Geoscience Australia, GPO Box 378, Canberra, ACT 2601, Australia}
}%

\jid{PASA}
\doi{10.1017/pas.\the\year.xxx}
\jyear{\the\year}

\usepackage{aas_macros}
\usepackage{hyperref} 
\hypersetup{colorlinks,citecolor=blue,linkcolor=blue,urlcolor=blue}

\begin{document}

\begin{frontmatter}
\maketitle

\begin{abstract}
We present the first data release (DR1) of the SkyMapper Southern Survey, a hemispheric survey carried out with the SkyMapper Telescope at Siding Spring Observatory in Australia. Here, we present the survey strategy, data processing, catalogue construction and database schema. The DR1 dataset includes over 66,000 images from the Shallow Survey component, covering an area of 17,200~deg$^2$ in all six SkyMapper passbands $uvgriz$, while the full area covered by any passband exceeds 20,000 deg$^2$. The catalogues contain over 285 million unique astrophysical objects, complete to roughly 18~mag in all bands. We compare our $griz$ point-source photometry with PanSTARRS1 DR1 and note an RMS scatter of 2\%. The internal reproducibility of SkyMapper photometry is on the order of 1\%. Astrometric precision is better than 0.2~arcsec based on comparison with Gaia DR1. We describe the end-user database,
through which data are presented to the world community, and provide some illustrative science queries.
\end{abstract}

\end{frontmatter}

\section{INTRODUCTION }
Maps of the sky have been produced throughout the cultural history of mankind. They represented the patterns in the sky, long believed to be static, for navigation and time-keeping; they provided a reference for measuring motions and allowed the identification of transient objects. Over time we came to understand our place in the Milky Way and the Universe at large, and discovered new phenomena such as novae and supernovae. From motions of celestial bodies we inferred the form and parameters of fundamental laws in physics such as the law of gravity and derived the existence of dark matter. 

In modern times, digital surveys such as the Sloan Digital Sky Survey \citep[SDSS;][]{York00} have revolutionised the speed and ease of making new discoveries. Instrumental limitations imply that surveys need to strike a balance between depth and areal coverage. The SDSS, e.g., imaged $\sim$70\% of the Northern hemisphere sky. In contrast, the SkyMapper Southern Survey presented here covers the entire Southern hemisphere. Many surveys carried out from space-based telescopes aim for all-sky coverage, e.g. the Gaia survey \citep{Gaia16a, Gaia16b}, the survey by the Widefield Infrared Survey Explorer \citep[WISE;][]{WISE}, the X-ray surveys from the Roentgen Satellite \citep[ROSAT;][]{ROSAT} and the eROSITA mission \citep{eROSITA}. These missions do not replace but complement each other by observing the sky at different wavelengths, which allows us to see different physical components of the matter in the Universe; or they have different specialities such as Gaia's high-precision astrometry, and they observe the sky at different times thus revealing variable phenomena at different stages.

Currently, a complete multi-colour inventory of the Southern sky to a depth of $g,r\approx 22$, similar to the SDSS in the Northern sky, is missing, and one aim of the SkyMapper project is to fill this gap. However, SkyMapper is not a Southern copy of the SDSS in the North, in the way in which e.g. the 2 Micron All-Sky Survey \citep[2MASS;][]{2MASS} used two telescopes in two hemispheres to create a consistent all-sky dataset from the ground. The SDSS in the North was especially successful in revealing new knowledge about distant galaxies, which should appear similar in the South given we live in an approximately isotropic Universe. Instead, SkyMapper aims for new discovery space beyond just repeating the SDSS in another hemisphere:

\begin{enumerate}

\item One focus of SkyMapper is studying the stars in our Milky Way, where the hemispheres are very different from each other. The Southern sky includes the bulge and centre of our Milky Way, as well as our two largest satellite galaxies, the Magellanic Clouds. Currently, we expect that the oldest stars in the Milky Way will be found in the bulge, and SkyMapper has indeed found the oldest currently known examples there \citep{Howes15}.

\item The SkyMapper filters differ from SDSS filters mainly in splitting the SDSS $u$-band into two filters: while the SDSS filter has $(\lambda_{\rm cen}/{\rm FWHM})=(358{\rm nm}/55{\rm nm}) $, SkyMapper has a violet $v$-band (384/28) and a more ultraviolet $u$-band (349/42). This choice of filters adds direct photometric sensitivity to the surface gravity of stars via the strength of the Hydrogen Balmer break driving changes in the $u-v$ colour, as well as to their metallicity via metal lines affecting the $v-g$ colour. By exploiting this information, SkyMapper has found the most chemically pristine star currently known \citep{Keller14}. We note that the other filters, $griz$, differ from their SDSS cousins in central wavelength and width by up to 40~nm (see Table~\ref{filtab}, and Fig.~\ref{SM_colcorrs} for colour terms).

\item Southern galaxies are of special interest when their distances are below e.g. 500~Mpc, as the distinct large-scale structure creates environments that differ from the ones in the North. While the Northern hemisphere has the iconic Virgo and Coma galaxy clusters, as well as the CfA2 and SDSS Great Walls \citep{GH89,GJ05}, the South distinguishes itself with the Shapley supercluster and the Southern extension to the Virgo Cluster.

\item As no ground-based telescope can observe the entire sky across two hemispheres, the synergy of combining data from different telescopes is limited by their location. Several cutting-edge wide-field radio telescopes are now being built in the Southern hemisphere, including the Australian SKA Pathfinder \citep[ASKAP;][]{Jo07,Jo08} and the Murchison Widefield Array \citep[MWA;][]{MWA} observing low radio frequencies, as well as the future Square Kilometre Array \citep[SKA;][]{SKA}. ASKAP and MWA are creating deep maps of the radio sky across more than the Southern hemisphere, and providing them with a homogeneous optical reference atlas is a major mission for SkyMapper.

\item More generally, cosmic dipoles have fundamental importance for understanding our place and motion within the Universe, and they may reveal new physics beyond the Standard Model. Without an all-sky view, the Cosmic Background Explorer \citep[COBE;][]{COBE} would have had a hard time measuring the motion of our Milky Way relative to the microwave background, which is a frame of reference on a scale large enough to not be dominated by its own peculiarities. Without an all-sky dataset, \citet{BW02} would not have corroborated this cosmic dipole by referring to distant radio galaxies. Current debates concern e.g. a cosmic dipole in the fine-structure constant of electrodynamics \citep[e.g.][]{Webb01,Murphy16}, and this work would greatly benefit from samples of quasi-stellar objects (QSOs) that cover most of the sky from pole to pole. SkyMapper will provide critical coverage of the Southern hemisphere to advance such studies.

\end{enumerate}

In the SDSS, spectroscopy was an integral part of the project, carried out at the same facility when the seeing was better matched to the task of feeding large fibres than to taking sharp images of the sky. From the start, SDSS planned to take a million spectra of bright galaxies as well as other objects of special interest, and over time the facility has been extended to allow dedicated spectroscopic surveys for several hundreds of thousands of stars, as well as over a million luminous red galaxies (LRGs) and nearly 300,000 QSOs \citep{SDSS3,SDSS_IV}.

SkyMapper, in contrast, was planned from the start as a pure imaging project, with periods of poor seeing dedicated to a transient survey \citep{Scalzo17} that finds supernovae by repeatedly imaging part of the sky, while using the main Southern Survey as a reference. Spectroscopy of a large number of objects was always going to be possible using the AAOmega spectrograph at the fibre-fed 2-degree Field facility \citep{2dF} at the 3.9-m Anglo-Australian Telescope, which can take spectra of nearly 400 objects simultaneously. Also, the detailed photometry of SkyMapper provides great power to preselect rare objects of interest, such as extremely metal-poor stars, and helps to minimise the number of spectra that might otherwise be required.

However, Australian expertise in multi-fibre instrumentation has now led to a new instrument for multi-object spectroscopy that greatly reduces the cost per spectrum in massive spectroscopic surveys. The new Taipan facility currently being commissioned at the UK Schmidt Telescope is expected to take up to a million spectra of stars and galaxies per year, and the eponymous Taipan galaxy survey will use SkyMapper for its source selection \citep{daCunha17}. One of its science goals is studying bulk flows in the local Universe from a peculiar velocity survey, and these can be much better constrained when data from both hemispheres are combined. 

While all these projects have their primary science drivers, we cannot overestimate their legacy value. As the datasets become available, the community of researchers finds new answers in established data repositories by asking new, creative questions that were not part of the initial project plan. In the long run, such legacy findings dominate the scientific output of most large projects that share their data with the broad community.

In this paper, we present the first Data Release of the SkyMapper Southern Survey (SMSS DR1)\footnote{http://skymapper.anu.edu.au}. This first major release presents data from the {\it Shallow Survey} covering nearly the whole Southern hemisphere. The master catalogue contains $\sim 285$ million unique astrophysical objects over 20,200~deg$^2$ of sky, while the photometry catalogue contains measurements of over 2.1 billion individual detections made in over 66,000 images. The catalogue is largely complete for objects of magnitude $\sim 18$ in all six filters across an area of 17,200~deg$^2$ and has saturation limits ranging from magnitude 8 to 9 depending on the filter and seeing. All SkyMapper magnitudes are in the AB system \citep{AB}. The specific release version described in this paper is DR1.1, which is released simultaneously to Australian astronomers and the world public as of December 2017. DR1.0 was released to the Australian community in June 2017, and used a different set of calibrator stars, which led to less well-calibrated photometry. Going to DR1.1, we were also able to identify and exclude a small fraction of poorly-calibrated images from the dataset.

The next release in preparation, DR2, will include images from the SkyMapper {\it Main Survey} that will reach 3-4~mag deeper, depending on the passband. Completing the Main Survey on the whole hemisphere will take until the year 2020. Even a decade from now, when data from the Large Synoptic Survey Telescope (LSST) will reach deeper into the Southern sky than any wide-field telescope before it, the SkyMapper Southern Survey will remain relevant, as it probes a different regime in brightness. All sources from this release will appear saturated and thus unmeasurable in LSST images, while the deeper SkyMapper Main Survey may assist in calibrating the LSST surveys, and adds several years to the time baseline in the study of variable phenomena with LSST. Finally, LSST lacks an equivalent to the $v$-band, such that SkyMapper will remain a source of important information on stellar parameters for brighter sources.

This paper is organised as follows: in Sect.~2 we give a short overview over the SkyMapper telescope and instrumentation, and summarise the main activities at the telescope; in Sect.~3 we describe the SkyMapper Southern Survey, while Sect.~4 details the procedures of the Science Data Pipeline (SDP), with a focus on the Shallow Survey and the resulting properties of the DR1 data; in Sect.~5 we discuss data quality and photometric comparisons; in Sect.~6 we explain data access methods as well as the database schema, and show a few example applications with database queries and results; finally, Sect.~7 provides an outlook to the future of the survey.

\section{S{\scriptsize ky}M{\scriptsize apper}}

SkyMapper is a purpose-built, optical wide-field survey facility located at Siding Spring Observatory near Coonabarabran in New South Wales, Australia, at the edge of the Warrumbungle National Park, which is Australia's first declared Dark Sky Park. It has a 1.35m primary mirror and uses a mosaic of 32 CCDs, each having $2048\times4096$ pixels with a scale of $\sim 0.5\arcsec$. The 268 million pixels cover a nearly-square field-of view measuring $2.4\times 2.3$ deg$^2$. After the commissioning phase ended in March 2014, observations for the Southern Survey started. The telescope is fully robotic and nightly survey operations are planned by autonomous scheduler software that can execute the entire survey without needing human intervention.

\subsection{Surveys and operations}

The public SkyMapper Southern Survey claims 50\% of the observing time on the SkyMapper Telescope, with a focus on good-seeing time, while bad-seeing time since 2015 has been used for the SkyMapper Transient Survey, a supernova survey described by \citet{Scalzo17}, and a fraction of time is kept available for third-party programs. The public Southern survey contains two components with different depth and saturation limits.

The {\it Shallow Survey} targets the entire Southern hemisphere with short exposures, ensuring full-hemisphere coverage early in the project and a bright saturation limit. Several dithered visits aim at fully covering the target area despite gaps in the CCD mosaic, and repeat observations to ascertain the static or transient nature of any detected sources. During each visit a six-colour sequence is observed within less than five minutes. While the first year of SkyMapper survey operations was dedicated mostly to the Shallow Survey, this now continues only in the brightest nights around full moon. The Shallow Survey is complete to nearly magnitude 18 in all six bands {\it uvgriz} and provides the calibration reference for the following Main Survey. 

\begin{figure*}
\begin{center}
\begin{minipage}{0.49\textwidth}
 \includegraphics[angle=270,width=\columnwidth]{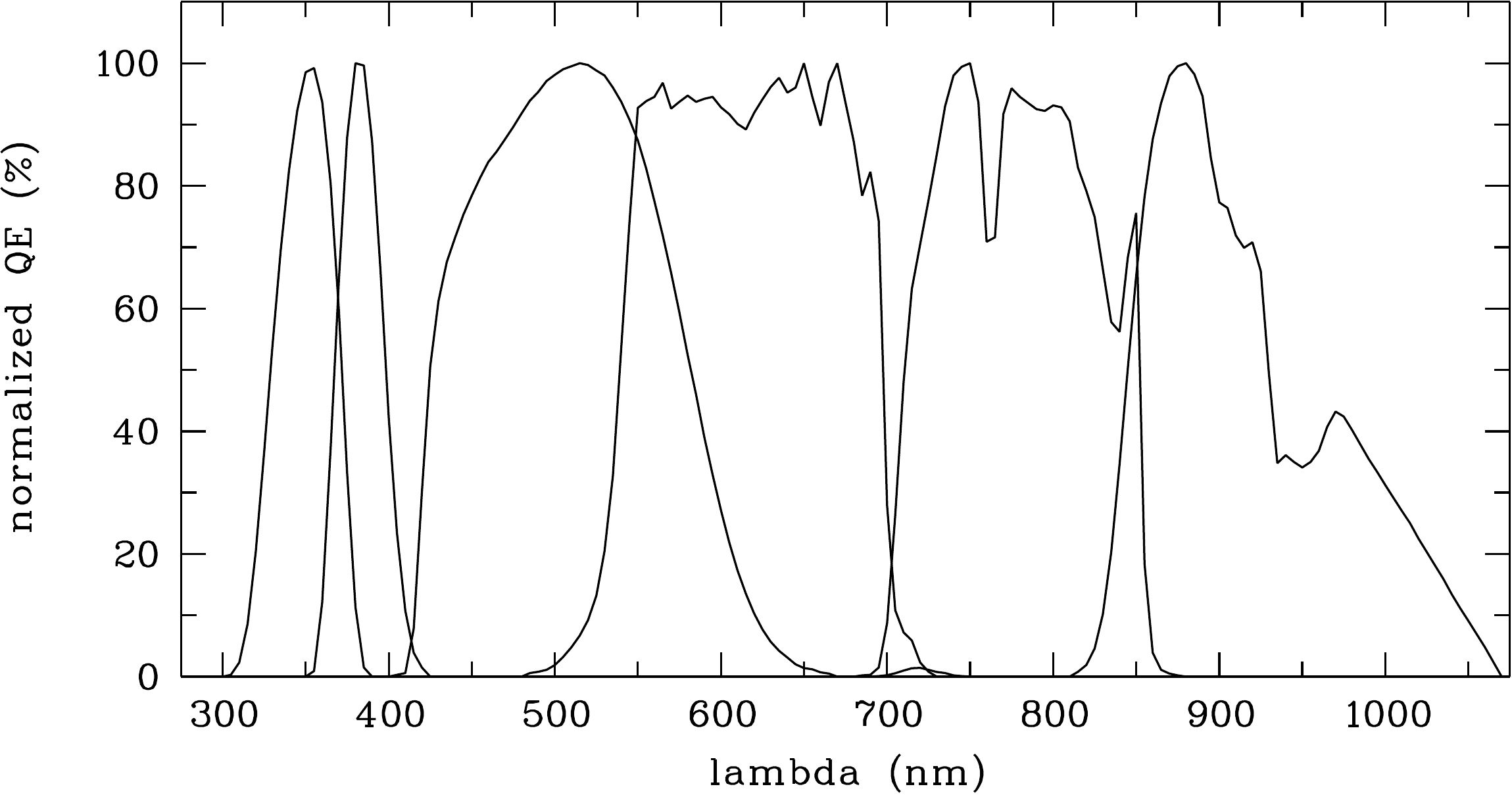} 
\end{minipage}
\hfill
\begin{minipage}{0.49\textwidth}
 \includegraphics[angle=270,width=\columnwidth]{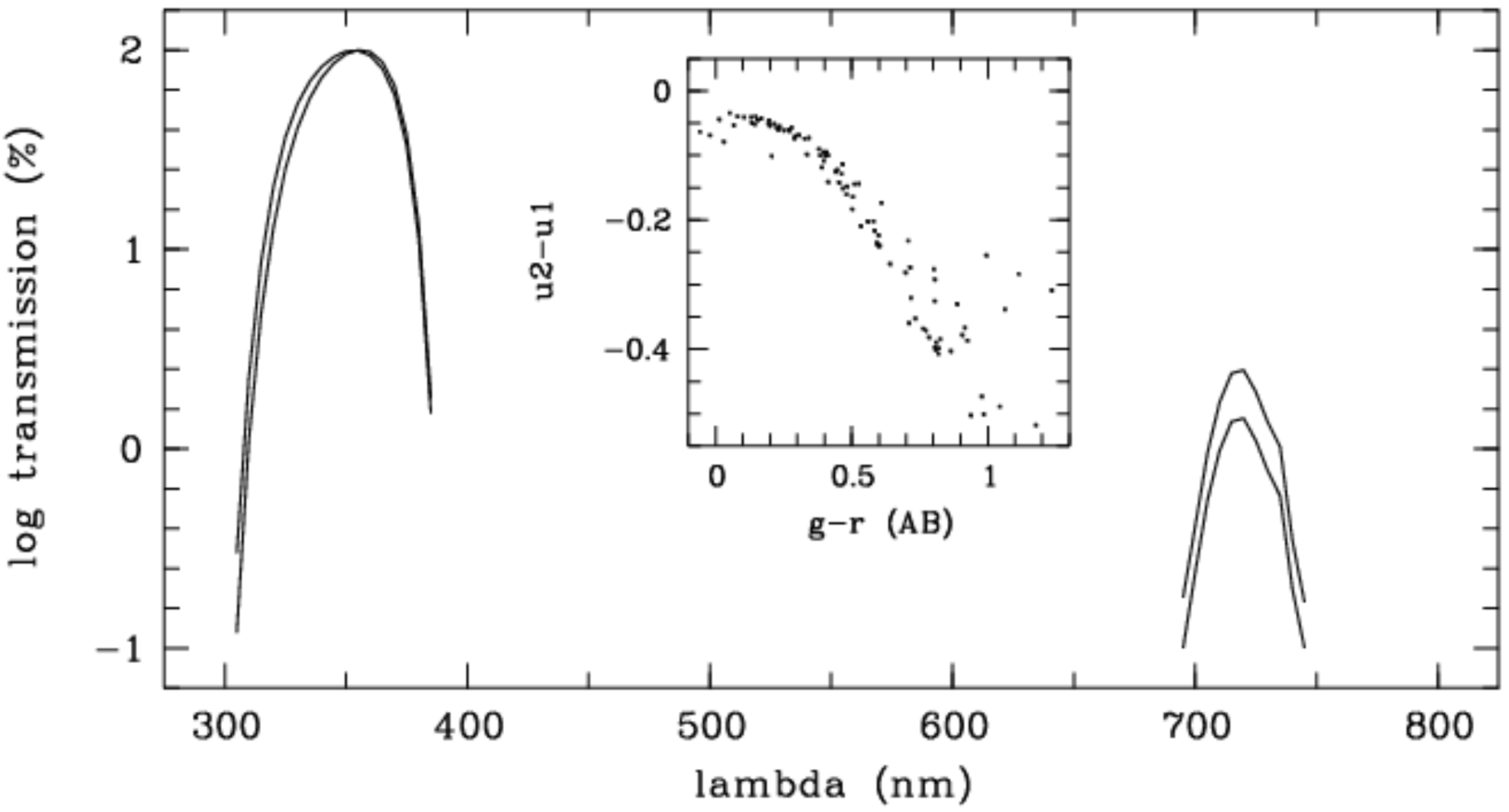} 
\end{minipage}
\caption{Normalised transmission curves for SkyMapper filters including atmosphere, telescope and detector from \citet{B11}. Note that SkyMapper has an ultraviolet $u$-band and a violet $v$-band, where SDSS has only a single $u$-band. Central wavelengths and filter widths of $griz$-filters vary up to 40~nm relative to their SDSS cousins. {\it Right:} $u$-band filter curve including atmosphere, telescope and detector, for airmass 1 and 2. The red leak at 700--750~nm wavelength increases relative to the main transmission band with airmass, because UV light is heavily absorbed by the Earth's atmosphere, while far-red light remains nearly unaffected. As a result, the measured $u$-band magnitudes of red stars increase with airmass as the relative contribution from the leak increases.}\label{filtercurves}
\end{center}
\end{figure*}

The {\it Main Survey} targets the same sky area in the same six bands, but aims to be complete to $g,r\approx 22$. In combination with the Shallow Survey, the entire SkyMapper Southern Survey will provide calibrated $uvgriz$ photometry from magnitude 9 to 22. The Main Survey includes two six-colour sequences, each taken within a 20-min interval, as well as additional visits to collect pairs of $gr$ and $iz$ images. The $gr$ pairs are acquired under dark and grey sky, while the $iz$ pairs are primarily collected in the period between astronomical and nautical twilight.

SkyMapper's regular operations model includes daily calibration procedures including bias frames, twilight flatfields and several visits to dedicated SkyMapper standard fields during the night. Eight such standard fields have been defined, each of which includes a {\it Hubble Space Telescope} (HST) spectrophotometric standard star. These standard field observations have not been used for DR1, but will be utilised in future releases. A paper with more complete technical and procedural references is in preparation (Onken et al. 2018).

\begin{table*}
\caption{Properties of SkyMapper DR1 imaging data. $10\sigma$-limits are quoted for the average object in the master catalogue. The PSF FWHM, zeropoint and background are median values among the 66,840 individual exposures. Most $uvgri$ images are read-noise limited in the Shallow Survey. $R_{\rm band}$ is a reddening coefficient from a \citet{F99} law with $R_V=3.1$.}
\label{filtab}      
\centering          
\begin{tabular}{l|lccccccrr}
\hline\hline       
Filter & $\lambda_{\rm cen}$/$\Delta \lambda$ & $R_{\rm band}$ & $t_{\rm exp}$ & $N_{\rm image}$ & $10\sigma$-limit & FWHM & zeropoint & bkgd & bkgd limit  \\ 
	   & (nm) & & (s) & & (ABmag) & (arcsec) & (ABmag) & (counts) & (counts) \\
\hline
$u$ & 349/42  & 4.294 & 40 & 10,909 & 17.9 & $3.1$ & 25.47 &  13 &  300  \\ 
$v$ & 384/28  & 4.026 & 20 & 10,779 & 17.7 & $2.9$ & 24.88 &   8 &  300  \\ 
$g$ & 510/156 & 2.986 &  5 & 11,872 & 18.0 & $2.6$ & 25.68 &  30 & 1000  \\ 
$r$ & 617/156 & 2.288 &  5 & 11,515 & 18.0 & $2.4$ & 25.46 &  31 & 1000  \\ 
$i$ & 779/140 & 1.588 & 10 & 10,698 & 18.0 & $2.3$ & 25.43 &  53 & 2500  \\ 
$z$ & 916/84  & 1.206 & 20 & 11,067 & 18.0 & $2.3$ & 25.47 & 128 & 2500  \\ 
\hline                  
\end{tabular}
\end{table*}

\subsection{Filters}

The SkyMapper filters are described in detail in \citet{B11}. The filters $u$, $v$, $g$ and $z$ all comprise coloured glass combinations; the $r$-filter was made by combining magnetron-sputtered short- and long-wave passband coatings; the $i$-filter is coloured glass plus a short-wave passband coating; the $z$-band long-wavelength cutoff is defined by the cutoff in the CCD sensitivity. All filters had two-layer anti-reflection coatings applied. The diversity in fabrication methods for the SkyMapper $griz$ filters helps explain why their passbands differ in width and placement from the SDSS $griz$ passbands.

After polishing, the $u$ and $v$ bandpass glasses were slightly thinner than anticipated. As a result, they have red leaks near 700 nm, around 1\% for $u$ and much less for $v$. These are evident in $u$-band observations of K and M stars (see Fig.~\ref{filtercurves}).

The transmission curves of the filters are shown in Fig.~\ref{filtercurves}; they were measured at many places across the aperture and averaged to provide the mean filter passband. The all-glass filters $u$, $v$, $g$ and $i$ have remarkably uniform wavelength transmission with position and with changing field angle. The interference-filter defined edges on the $r$- and $i$-filters, however, change slightly with angle. The average transmission was then multiplied with one airmass of atmospheric extinction and the mean QE response curve of the E2V CCDs. The CCD response curves and the interference coatings may change over the period of the survey and thus our calculated SkyMapper passbands may need adjustment. This investigation is beyond the scope of DR1 and will be assessed from observations of our spectrophotometric standard stars.

\begin{figure}
\begin{center}
\includegraphics[width=\columnwidth]{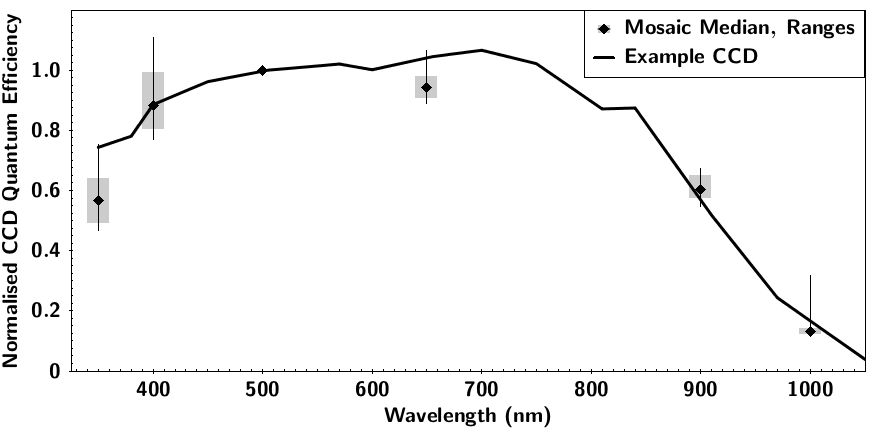}
\caption{Quantum efficiency (QE) of the CCDs in the mosaic of the SkyMapper camera. Points indicate the median values after normalising to a common level at 500 nm wavelength, the grey bars indicate the 16th-to-84th-percentile range at each wavelength, and the thin errors bars indicate the minimum and maximum values. A more detailed QE curve for a single typical CCD is shown with a line. Different short-wavelength sensitivities are evident, which will lead to subtle variations in the $uvg$ passbands.}\label{CCD_QEF}
\end{center}
\end{figure}

\subsection{Detector Properties}

SkyMapper's 32 detectors are arranged in four rows of eight CCDs, with horizontal spaces of $\sim 0.8$ arcmin between the mosaic columns, a small gap of $\sim 0.5$ arcmin between the central rows, and larger gaps of $\sim 3.2$ arcmin separating the two upper rows from each other and the two lower rows from each other. The read-out noise of the amplifiers varies broadly between 6 counts and 14 counts. In Shallow Survey exposures, fringes are only apparent in the $i$- and $z$-band. As their amplitude is below the read-out noise, we do not defringe any of the DR1 data released here.

Fig.~\ref{CCD_QEF} shows the efficiency curves of the 32 CCDs after normalising them at 500 nm wavelength. Variations in the blue sensitivity are already evident around 400 nm, which means that the $uvg$ passbands will differ from CCD to CCD, with corresponding subtle differences in the zeropoint calibration and colour terms. These variations are not considered in this data release, but will be included in a future release.

\begin{figure*}
\begin{center}
\includegraphics[width=2.1\columnwidth]{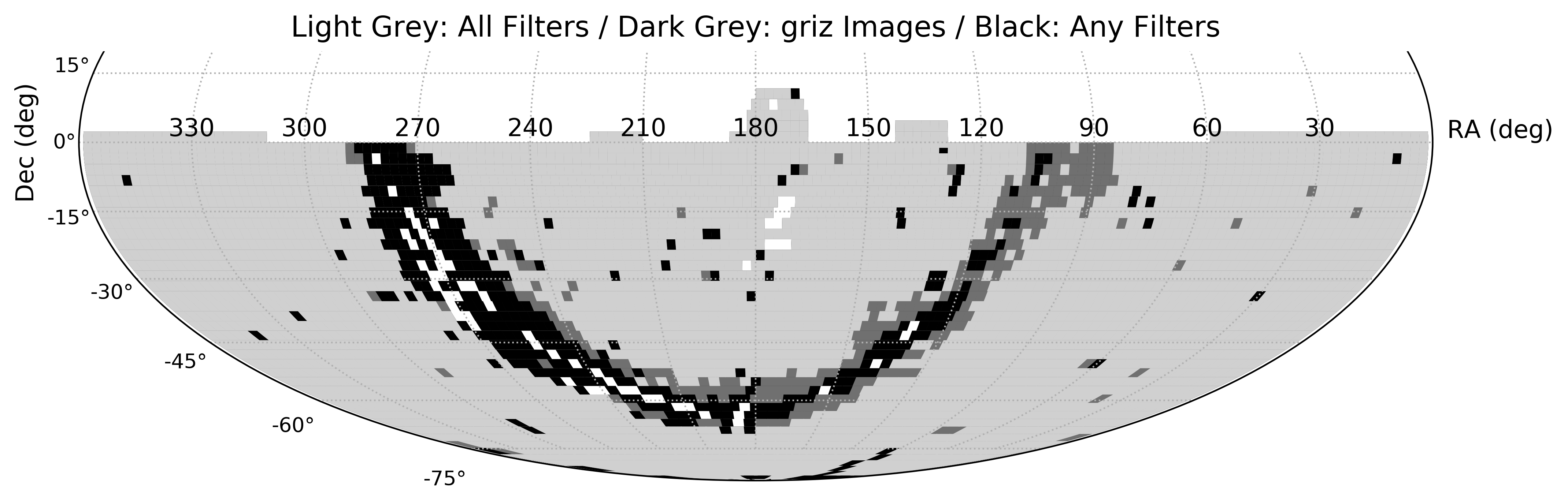}
\caption{Sky coverage of the SkyMapper Southern Survey Data Release 1 (SMSS DR1): Most of the Southern hemisphere has been covered and calibrated in good conditions in all six passbands (light grey), but some fields along the Galactic plane have only data for $griz$ in DR1 (dark grey), while fields without complete $griz$ coverage are shown in black and fields without data are left white. }\label{coverage2}
\end{center}
\end{figure*}

\section{The SkyMapper Shallow Survey}\label{SSS}

The observations for the SkyMapper Southern Survey started after the end of the commissioning period, on 15 March 2014. The first year of observations was mostly dedicated to the Shallow Survey, where three visits were planned per field. Since the end of year 1, further Shallow Survey images have been collected in bright time around full moon, and we plan to reach five visits per field by the end of regular survey operations in the year 2020. Visits to the same field are scheduled to occur on separate nights, but no special constraints are applied, so the time separation between of any pair of visits on a given field varies from 20 hours to over 500 nights; the median time between consecutive pairs is 10 nights.

Between 29 April and 29 May 2014 a fault in the filter mechanism forced us to observe with only one filter per night, and in this period SkyMapper observed an intense calibration program instead of the Shallow Survey. On 24 September 2015 the helium compressor of the detector cooling system failed terminally. Since operations restarted with a new compressor two months later, the images show localised but strong electronic interference, which requires additional processing that is not yet available. 

Hence, DR1 is focused on Shallow Survey data obtained in the first-compressor era 2014-15. During this period several changes occurred that affect the image and calibration quality in DR1 in subtle ways, including a change in flatfield strategy implemented in November 2014 (see Sect.~\ref{flat_creation}), and an early period until 1 July 2014 that is affected by certain detector artefacts (see Sect.~\ref{sec_tearing}). 

DR1 includes 66,840 images after deselecting images of insufficient quality. The typical image properties are summarized in Table 1. Our quality cuts include:
\begin{enumerate}
 \item upper limits to the PSF FWHM ($5\arcsec$ for $griz$ and $6\arcsec$ for $uv$) and elongation ($<1.4$);
 \item lower limit to the image zeropoints (total instrument efficiency including atmospheric transmission) of roughly 24.5~mag, except 23.8~mag for $v$-band;
 \item upper limit to the rms among the approximate zeropoints derived from the calibrator stars within an image (only approximate, as these were applied before the DR1.1 recalibration), ranging from $\sim 0.1$ for $z$-band to $\sim 0.2$~mag for $u$-band;
 \item upper limits on the background level to limit sky noise;
 \item the number of well-measured calibrator stars in a frame needs to be at least 10 before a final clipping of outlier zeropoints. 
\end{enumerate}

The median FWHM of the PSF among all DR1 images ranges from $2.3\arcsec$ to $3.1\arcsec$ from $z$-band to $u$-band, and the median PSF elongation is 1.12 independent of filter. Most images of the Shallow Survey are read-noise limited as the median sky background is less than 100 counts, except for the $z$-band images that are mostly background-limited. The median airmass of observations is 1.11, but a tail to airmass 2 is unavoidable given that the sky coverage includes the South Celestial Pole. 

The image zeropoints are largely around 25.5~mag, with the exception of the shallower $v$-band. The zeropoints show long-term drifts towards lower efficiencies until the mirrors were cleaned on 5 May 2015, resulting in an improvement by $\sim$0.35 mag. Most image zeropoints for a given filter scatter within less than 0.1~mag rms at a given calendar epoch, but some nights with bad weather produce tails with higher atmospheric extinction; the fraction of images with zeropoints that are shallower by over 0.3~mag than the median zeropoint noted in Tab.~\ref{filtab} increases monotonically with light frequency from 5\% in $z$-band to 12\% in $u$-band.

Gaps between the CCDs in the detector mosaic imply that some sky coordinates are imaged less frequently than suggested by the number of field visits. Overlaps between neighbouring fields, however, mean that areas close to the field perimeters are observed many more times. The objects in the master table have visit numbers ranging mostly from 0 to 5 depending on filter, field and coordinate, but the tail reaches up to 17 visits in $g$ and $r$-band, and, exceptionally, 35 visits in $u$-band. In $z$-band, e.g., the mean number of visits per object is 2.31, and about 1\% of objects have more than five visits. Overall, the mean number of visits per object ranges from 2.0 in $v$-band to 2.8 in $g$-band. 

Furthermore, objects might fall on bad pixels, which is ignored when computing the number of visits. Instead, these objects appear in the catalogues with flags indicating the presence of bad pixels. Every image has an associated mask image that combines globally bad pixels with frame-specific bad pixels. Globally bad pixels are identified by a combination of algorithmic and manual processes and account for 0.31\% of the pixels in each image. Frame-specific bad pixels include e.g. pixels affected by cross-talk from saturated pixels in an associated amplifier (see Sec.~\ref{sec_det_cross}). 

The DR1 coverage map of the Southern sky (Fig.~\ref{coverage2}) shows some areas with missing data: these include (1) regions that were not yet visited in good-quality conditions, which is the most common reason off the Galactic plane, and (2) regions that lack well-measured and reliable calibrator stars, which is the most common reason close to the Galactic plane. In DR1, photometry is heavily compromised by stellar crowding, lowering the number of cleanly measured (isolated) stars in dense star fields. This problem will be overcome in future releases. Also, our photometric calibration source (see Sec.~\ref{ZPcalib}) contains areas where star colours are inconsistent with the expected stellar locus by more than 0.2~mag and those areas are not included in DR1.1, although they were part of DR1.0.

\begin{figure}
\begin{center}
\includegraphics[width=\columnwidth]{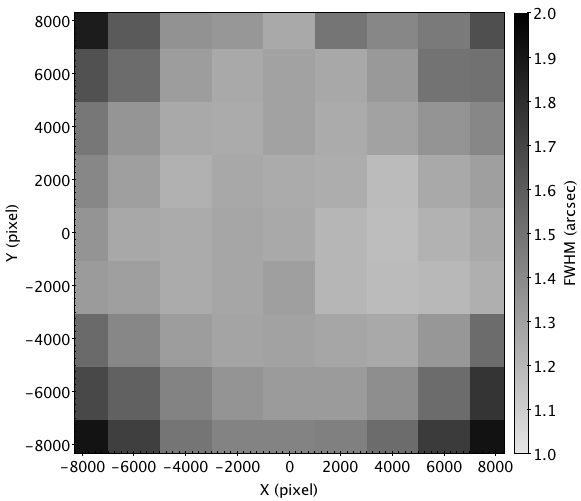}
\caption{Map of the PSF FWHM across an $i$-band image taken in good seeing: due to a curved focal plane the FWHM ranges from $1.2\arcsec$ to $1.8\arcsec$.}\label{PSF_FWHM}
\end{center}
\end{figure}

Within images, we note significant PSF variations caused by a curved focal plane, so they appear most pronounced in good seeing. We generally attempt to minimise the PSF variations by putting best focus not into a point in the centre, but into a ring. The PSF then deteriorates mildly towards the centre and moderately towards the corners. However, due to focus drifts at the telescope, the ring of best focus will breathe and cause varying patterns of deterioration. Fig.~\ref{PSF_FWHM} shows an example of an $i$-band image obtained in good seeing, where the PSF FWHM varies from $1.2\arcsec$ to $1.8\arcsec$. In other images we see e.g. variations from $2.8\arcsec$ to $3.5\arcsec$, but a thorough investigation is beyond the scope of this paper. As we describe in Sect.~\ref{apercorr}, we compensate for PSF variations in the point-source photometry by constructing a 1D-PSF photometry estimate based on a sequence of aperture magnitudes and their trends across any individual image.

In DR1, we do not yet include any illumination correction to remove deviations of the measured flatfield from the true sensitivity pattern. A first analysis of illumination correction maps suggests that these correction are below $\pm1$\% on 90\% of the mosaic area in all six filters. However, these corrections can reach up to 5\% in the very corners of the image. As a result, a small fraction of objects situated close to field corners could be imaged under varying illumination conditions that are presently not accounted for. This effect will be corrected in future releases. Some images contained in the release are affected by dome vignetting, which causes a moderately high rms of the zeropoint estimate due to the changing collecting area across the field-of-view.

\section{Science Data Pipeline}

\begin{figure}
\begin{center}
\includegraphics[width=\columnwidth]{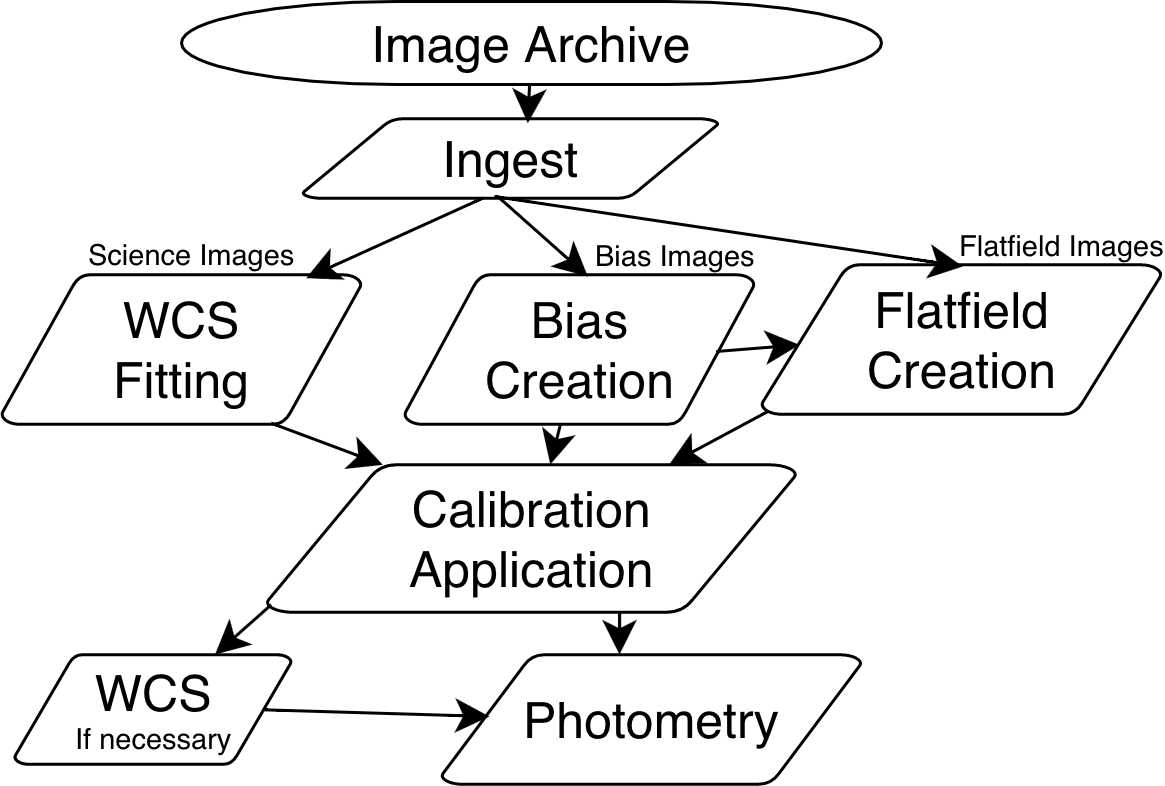}
\caption{Flow chart of the steps in the Science Data Pipeline.}\label{flowchart}
\end{center}
\end{figure}

The images taken by SkyMapper are processed through the Science Data Pipeline (SDP), which utilises a PostgreSQL\footnote{http://www.postgresql.org} database to semi-autonomously oversee the ingestion of each mosaic image, control the creation of the required calibration frames, produce reduced images, and perform photometric measurements and calibration \citep{ADASS_Luvaul,ADASS_Wolf}. The flow of the SDP is outlined in Figure~\ref{flowchart} and the sections below. 

\subsection{Ingest}

Images from the telescope are transferred each morning via ethernet to long-term storage at the National Computational Infrastructure (NCI) located on the ANU campus. When a night is ready to be processed by the SDP, the data is moved from the NCI image archive to a local disk, and the individual images are then ingested into the SDP, which reorganises the 64 extensions of the FITS file (one per amplifier) into 32 separate FITS images (one per CCD). Preliminary quality assurance (QA) checks are performed to eliminate bad images from further processing. The initial processing during the ingest phase includes: suppression of interference noise; identification of pixels subject to saturation or blooming; overscan subtraction; correction for cross-talk; creation of image masks; and flipping the image axes to restore the on-sky orientation.

\subsubsection{Interference Noise}

\begin{figure}
\begin{center}
\includegraphics[width=\columnwidth]{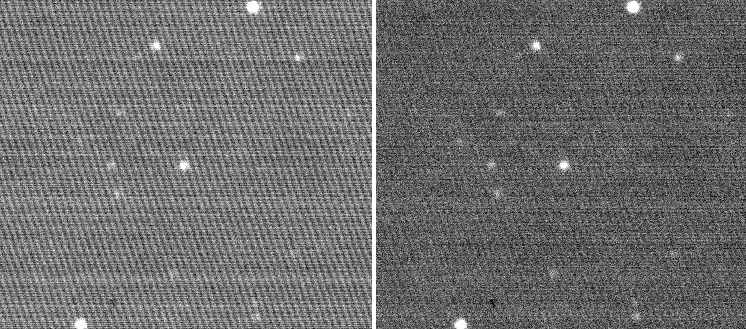}
\caption{{\it Left:} Interference noise that appears intermittently in SkyMapper images, with amplitudes and row-to-row phase shifts that vary within a single readout. The wavelength along each row is between 6 and 7 pixels. We obtain a sinusoidal fit to each row, which is then subtracted to remove the interference. {\it Right:} The same image after removal of the fit to the noise. Residual row-to-row variations in the bias are removed by a separate step.}\label{fig_interference}
\end{center}
\end{figure}

The SkyMapper detectors are subject to an intermittent, high-frequency source of interference, which imposes a strong sinusoidal pattern on the images with a wavelength between 6 and 7 pixels. The phase and amplitude can vary from row to row in an image, so each row read out by each amplifier is corrected individually by fitting a sine curve. In rare cases, the procedure does not adequately remove the pattern from a given row. The mean amplitude of the subtracted sine curve was 4.4 counts. An example of the interference noise and its removal are shown in Figure \ref{fig_interference}. 

\subsubsection{Saturated and Bloomed Pixel Identification}

For all image types except bias frames, we run Source Extractor \cite[version 2.19.5;][]{BA96} with a configuration designed to pick up heavily saturated sources: detection threshold of 58,000 counts and minimum area of 25 pixels. This identifies very bright sources as well as the surrounding pixels that are affected by blooming, and these are treated specially in the cross-talk correction (Sect.~\ref{sec_det_cross}).

\subsubsection{Overscan Correction}

Because the bias level has less structure in the post-scan regions (towards the centre of each CCD) than the pre-scan regions (Sec~\ref{sec_det_bias}), the post-scan level in each row is used for the overscan correction. Taking the last 45 pixels in the 50-pixel post-scan region, the set of pixels is 3$\sigma$-clipped until convergence, and the median of the resulting distribution is subtracted from the row.

\subsubsection{Amplifier cross-talk and ringing}\label{sec_det_cross}

\begin{figure}
\begin{center}
\includegraphics[width=\columnwidth]{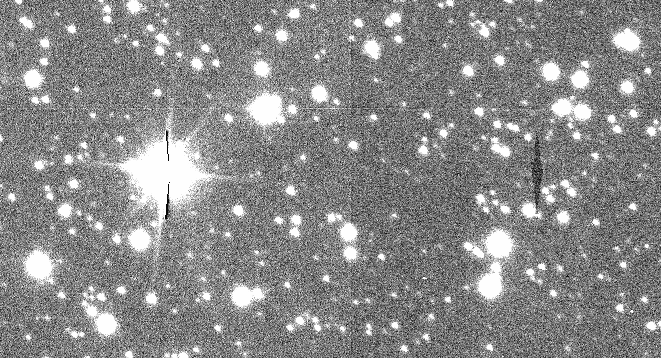}
\caption{Example image showing cross-talk between amplifiers, visible as depressed counts opposite the bright star on the left, and amplifier ringing, visible as dark regions immediately adjacent to the pixel blooming associated with the bright star on the left.}\label{fig_crosstalk}
\end{center}
\end{figure}

The SkyMapper mosaic is subject to electronic cross-talk between its 64 amplifiers. The most severe cross-talk, and the only one treated by the current version of the SDP, occurs between adjacent amplifiers on the same CCD, with a fractional amplitude of $\sim 5\times10^{-4}$. However, in the presence of fully saturated pixels, other amplifiers can display cross-talk artefacts (switching between negative and positive counts on alternative amplifiers) with fractional amplitudes as high as $1\times10^{-4}$ or about 7 counts. Fully saturated pixels also induce amplifier ringing, where the next non-saturated pixel read out has 0 counts and the subsequent two pixels have severely suppressed count levels. Figure~\ref{fig_crosstalk} shows examples of these behaviours.

\begin{figure}
\begin{center}
\includegraphics[width=\columnwidth]{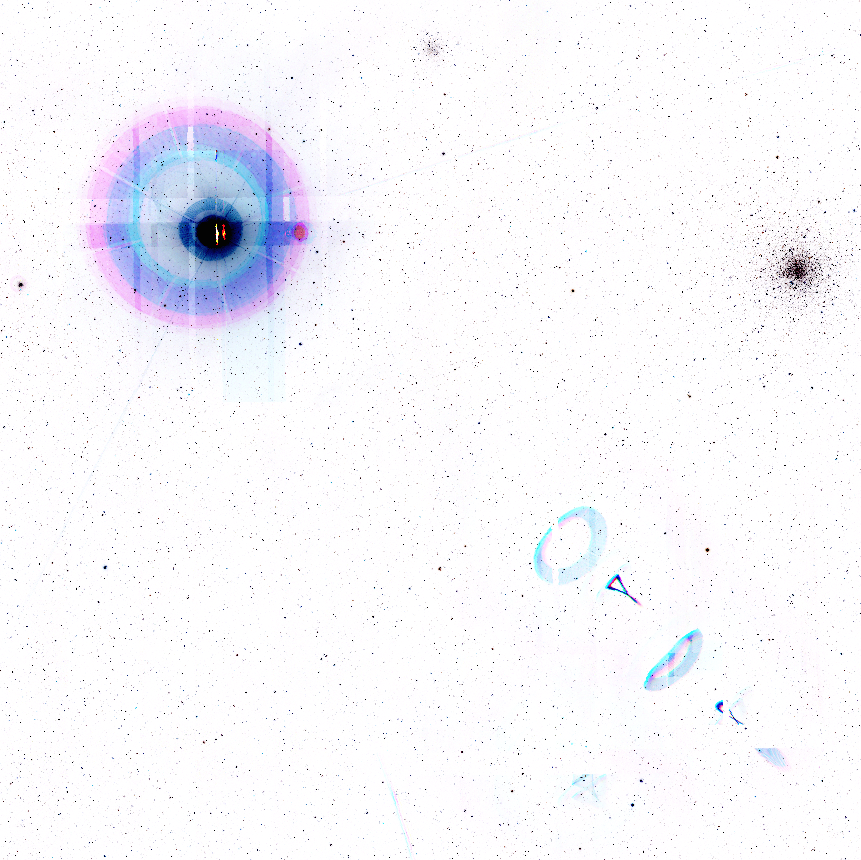}
\caption{Optical reflections from very bright stars (here, Antares) are not removed from images in DR1. This image is an inverted-colour image from the SkyMapper SkyViewer.}\label{opt_reflections}
\end{center}
\end{figure}
Cross-talk between the two amplifiers on the same CCD is corrected by subtracting a second-order polynomial in the count level, with an additional subtractive component for the counterpart of a fully saturated pixel. The correction is made in a single pass through each CCD, beginning with the left-hand amplifier and then correcting the right-hand amplifier.\footnote{This occurs before the image axes are flipped, so the ordering is reversed with respect to the reduced images in the release.} 

The pre- and post-scan pixels are then trimmed from both amplifiers of a given CCD, and the two amplifiers are stitched together into a single FITS image. Finally, the images and masks are flipped in the x-direction to restore the on-sky orientation.

\subsubsection{Detector bias}\label{sec_det_bias}

The SkyMapper detectors exhibit no significant fixed-pattern bias structure, but there are two kinds of row-oriented, time-variable bias features that are seen. The first is a slow change in the bias level along the row, with the spatial scale of the variation increasing from the pre-scan towards the post-scan in each amplifier. The variations can be well modelled by a Principal Components Analysis described in Sect.~\ref{bias_creation}. 

The second kind of bias features are known as "fingers", and consist of regions of suppressed bias level near the edges of the frame. Most of the SkyMapper CCDs show some amount of this behaviour, but in a small number of CCDs, a fraction of the rows (which change from readout to readout) have the suppressed bias level extending from the pre-scan region into the on-sky portion of the detector. The amplitude of the bias offset is typically 20 counts, with a distribution that extends to $\sim40$ counts, and the bias makes a sharp transition back to the nominal level for the rest of the row. These features remain uncorrected in DR1.

\begin{table}
\caption{Meanings of image mask flags.}
\centering
\begin{tabular}{@{}ll@{}}
\hline\hline
Bit & Description \\
\hline%
0 & Hot/Cold pixel from Global Bad Pixel Mask\\
1 & Not used\\
2 & Saturated (>55k counts in original frame)\\
3 & Affected by detector cross-talk\\
4 & Not used\\
5 & Cosmic ray\\
\hline\hline
\end{tabular}
\label{tab_maskbits}
\end{table}

\subsubsection{Mask Creation}

Image-specific masking augments the global bad pixel mask. We identify saturated pixels using a threshold of 55,000 counts. The fully saturated pixels and the three neighbouring pixels subject to amplifier ringing are all masked with a saturation flag. Pixels subject to cross-talk from saturated pixels in the other amplifier on the same CCD are masked with a separate flag. Finally, as described in Sect.~\ref{sec_otzffi}, pixels in which cosmic rays have been identified and corrected are flagged. The various image-mask flags are described in Tab.~\ref{tab_maskbits}.

Very bright stars leave some strong and curiously shaped optical reflections in the images, which have not been removed (see Fig.~\ref{opt_reflections} for an example).

\subsection{Astrometry}\label{sec_astrom}

The astrometric solutions for the survey images are tied to the fourth U.S. Naval Observatory CCD Astrograph Catalog \cite[UCAC4;][]{2013AJ....145...44Z}. The 32 CCDs are each run independently through the {\sc solve-field} program from the Astrometry.net software suite \cite[version 0.43;][]{2010AJ....139.1782L}, using a set of UCAC4 index files, to generate tangent-plane World Coordinate System (WCS) solutions. Each CCD for which solve-field succeeds is then run through Source Extractor to derive ($x$, $y$) pixel positions and RA, Dec values for each source, as well as to calculate the median seeing and elongation values for the CCD (using Source Extractor parameters FWHM\_IMAGE and ELONGATION). For each source in the image, the closest-matching RA, Dec position from the UCAC4 catalogue is taken as the true position\footnote{We do not update the UCAC4 positions to the observed epoch based on the UCAC4 proper motion measurements, but do not expect this to contribute significant errors to our final astrometic solutions.}, and we then fit a Zenithal Polynomial (ZPN) projection \cite[see][]{2002A&A...395.1077C} to the pixel+coordinate sets using a fixed curvature term of PV2\_3 = 3.3, which was determined by an independent fit to a well sampled mosaic image. The ZPN solution provides a good fit to the radial distortions in the SkyMapper focal plane, but the ZPN formalism is not understood by several of the software tools utilised by the SDP. Hence, the ZPN solution is then transformed into a TPV projection (tangent plane with generalised polynomial distortion)\footnote{https://fits.gsfc.nasa.gov/registry/tpvwcs/tpv.html}, which can be understood by the software packages from AstrOmatic\footnote{http://www.astromatic.net/} (e.g., SExtractor, SWarp, and PSFEx). 

After all 32 CCDs have been analysed, two QA tests are performed: the plate scale near the middle of each CCD is required to be between $0.487\arcsec$ and $0.507\arcsec$ ($0.483\arcsec$ and $0.507\arcsec$ for the four corner CCDs); and the offset from the middle of the CCD to the middle of one of the central CCDs is required to be within $10\arcsec$ of the nominal value. Later, a third test is performed on the WCS solutions, requiring that the corner-to-corner lengths of each CCD are within $2\arcsec$ of a reference value, where the reference is dependent on the CCD position within the mosaic, the filter used in the observation, and the fractional Modified Julian Date (MJD). The dependence on the fractional MJD, which contributes $0.5\arcsec$ of variation across the observed fractional MJD range, is thought to reflect temperature gradients in the Corrector Lens Assembly, which evolve as a night progresses. This final test ensures that the WCS solutions do not extrapolate to unrealistic representations of the sky in regions where there may be too few stars to constrain the fit.

Overall, 98.6\% of mosaic images used in DR1 yield valid WCS solutions for all 32 CCDs, and amongst the few images that yield between 1 and 31 valid WCS solutions, the mean number is over 28. The success rate is lowest in the $u$- and $v$-filter with 96.7\% each, because the density of stars bright enough to match to UCAC4 is the lowest. Whenever a valid WCS solution is not available for a CCD, it has further calibrations applied as normal, after which a second attempt is made to obtain a WCS solution.

\begin{figure*}
\begin{center}
\includegraphics[width=\textwidth]{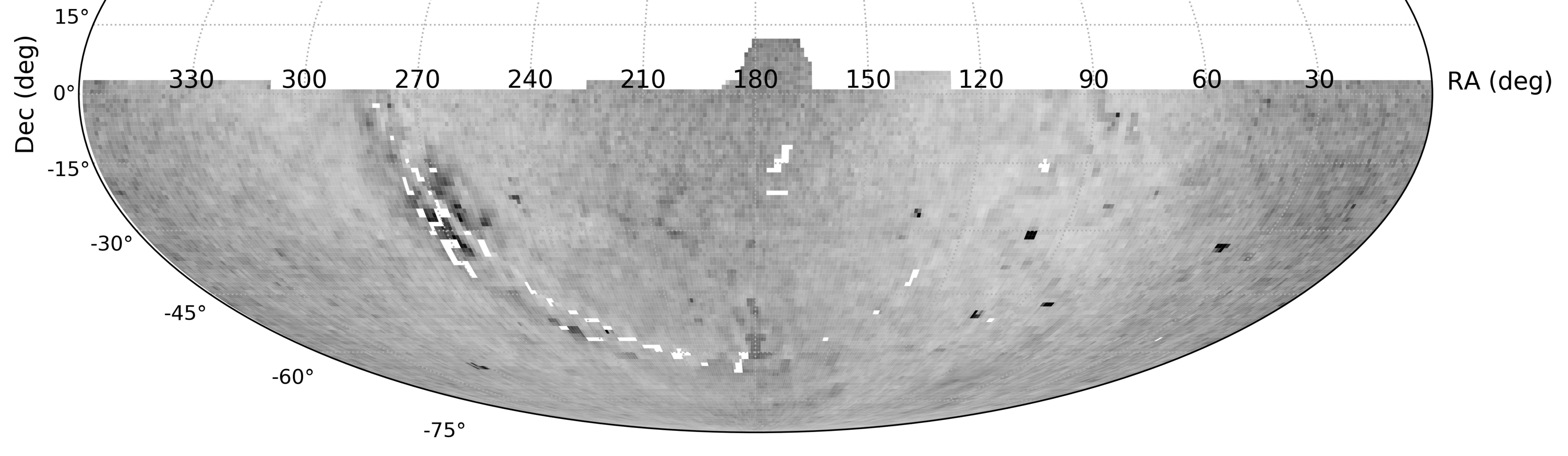}
\caption{Median positional offset between the DR1 sources and the closest match from the Gaia DR1 catalogue, as a function of Right Ascension and Declination. The grayscale ranges from 0 (white) to 0.5~arcsec (black). The overall median offset is 0.16~arcsec.}\label{GaiaOffsetMap}
\end{center}
\end{figure*}

As an indicator of the quality of our WCS solutions in DR1, we show the median offset distances between the sources in our DR1 master table and the nearest source in Gaia DR1 \citep{2016A&A...595A...4L} as a function of RA and Dec in Figure \ref{GaiaOffsetMap}. The overall median offset (restricted to distances less than 10$\arcsec$) is $0.16\arcsec$ and the $[10\%,90\%]$ range is $[0.06\arcsec, 0.45\arcsec]$. For cleanly detected objects with $r$-band magnitudes between 9 and 14~mag, the median offset is $0.12\arcsec$, or about 1/20th of the typical PSF FWHM.

\subsection{Calibration Data}

The individual bias and flatfield exposures are combined into a series of master frames with variable durations of validity. Hard boundaries on the input frames considered for each master calibration frame are imposed at each instance of detector warming (when we find subsequent rapid evolution of the flatfield shape), as well as at any other significant change to the image properties, such as modification of detector voltages (Sect.~\ref{sec_tearing}), changes to flatfield position angles, or cleaning of the telescope optics.

\subsubsection{Bias Creation}\label{bias_creation}

The typical observing night includes the acquisition of 10 bias exposures, which are used to generate the master bias frame valid for the night. In cases when fewer than 10 biases were obtained, the adjacent nights are considered. In practice, all master bias frames have at least 10 inputs, and for most nights, all inputs come from the night in question. The maximum range of input nights is 8.

From the input biases selected, the master bias frame is created as the mean image while rejecting outliers that are 50 counts above or below median for each CCD. Pixels from the global bad pixel mask are set to 0 counts. The master bias is then subtracted from each of the input biases, which then are used to generate the night-specific Principal Components (PCs) for the residual bias shape along each row \citep[using the {\it scikit-learn} Python module;][]{sklearn}. We treat the two halves of the CCD separately to account for the differing behaviour between amplifiers, and for each amplifier we compute the mean and next 10 principal components of the per-row bias pattern. In addition, we derive limits for the allowable PC amplitudes when later fitting the science frames. The limits are set to trim off the most extreme 0.25\% of values from either end of the bias frame PC amplitude distribution.

\subsubsection{Flatfield Creation}\label{flat_creation}

The approach for considering potential inputs to a master flatfield differs from the bias frames in several respects. To minimise the effect of spurious variations in the observed flatfield images due to e.g. clouds, \emph{all} flatfield images within $+/-$ 10 nights are considered, except where that range encounters one of the hard boundaries between calibration windows. This results in a median range of 17 nights. The potential inputs are then trimmed of any inputs in which \emph{any} of the 32 CCDs exceeds certain tolerances in terms of the CCD count level relative to the mean of the central 8 CCDs\footnote{The central 8 CCDs (2 rows $\times$ 4 columns) comprise the middle 25\% of pixels from the mosaic, which demonstrate more stable behaviour with respect to the evolution of the flatfield shape.}. The tolerances are set to $+/-$ 10\% of the mean count ratio defined over a long time baseline and specific to each filter, position angle (PA), and twilight (evening/morning). The median number of inputs to the master flatfields ranges from 65 ($z$-band) to 85 ($v$-band).

The algorithm employed for flatfield creation then depends on whether the inputs were obtained with fixed PA ($0^\circ$; from the start of the survey through 19 Nov 2014) or with opposing-pair PAs. On 25 November 2014, we started taking flatfields in rotated pairs to capture and eliminate the gradient in sky brightness across our large field-of-view\footnote{For the period of 20-24 Nov 2014, a set of intermediate PAs was employed, but these did not provide satisfactory data, and the master flatfields from the beginning of the opposing-pair era have had their validity period extended backwards to cover those nights.}.

In the fixed-PA era, the input flats are first bias-corrected using the master bias frame.\footnote{No PCA bias correction is applied, as the row-bias variations are small compared to the noise in the individual flats.} For most master flatfields, the inputs are median-combined with 3$\sigma$-clipping, including scale factors defined as the inverses of the mean count levels in the central 8 CCDs of each input (to preserve the relative scaling of each CCD within the mosaic). For one master flat in $r$-band, there was only three input exposures available, and in that case we took the minimum value at each pixel as a way of rejecting any astronomical sources present in the input images.

For the opposing-pair era, the potential input frames are grouped by PA, and a similar tolerance-testing procedure is applied as in the fixed-PA era. Within each PA, the remaining inputs are median-combined after rescaling the counts by the mean of the central 8 CCDs. For the opposing PAs within each twilight, the PA-medians are then mean-combined with equal weighting. Finally, the twilight-means are combined in a weighted mean, where the weights are taken as the number of contributing input frames, resulting in the master flatfield.

In both the fixed-PA and opposing-pair eras, pixels from the global bad pixel mask are set to a value of 1.

\subsubsection{Detector "tearing"}\label{sec_tearing}

\begin{figure}
\begin{center}
\includegraphics[width=\columnwidth]{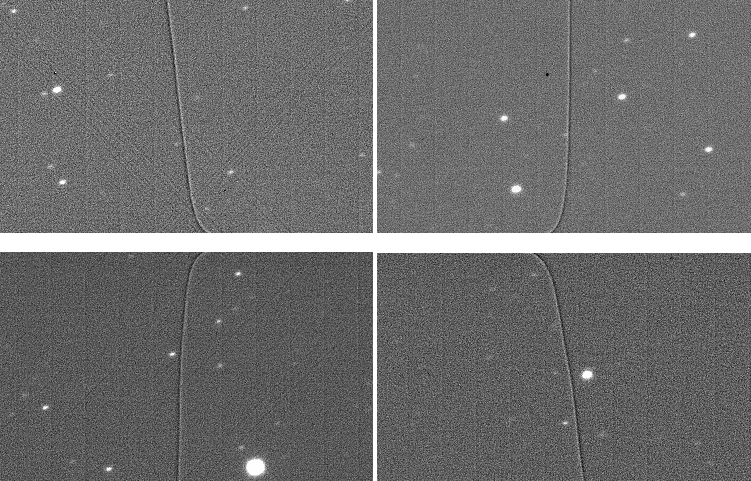}
\caption{Detector "tearing" on four amplifiers across four CCDs from an $r$-band flatfield image from UT 31 March 2014. The amplitude of the features is approximately 5\% of the nominal count level. The position of the features varied with amplifier and with each image. Modifications to the detector bias voltages eliminated the features from July 2014 onward.}\label{detectorlines1}
\end{center}
\end{figure}

In early 2014, it was discovered that all 32 of the CCDs exhibited curvilinear features in which approximately 5\% of the counts appeared to be shifted to neighbouring columns. An example is shown in Figure \ref{detectorlines1}. The position of these features differed from amplifier to amplifier and changed position from image to image, and hence we leave them uncorrected in DR1. The features were eventually identified as being instances of "tearing" \cite[cf.][]{Doherty13,2014JInst...9C4027R} and were traced to the detector bias voltage settings being used at the time. From July 2014, an updated configuration, modelled on the settings used by the European Southern Observatory, eliminated the features from subsequent SkyMapper images. However, the relative-strength nature of the tearing means that in DR1 the features may be visible in flatfields with few input frames or in high-background images taken prior to July 2014.

\subsection{Application of Calibrations}\label{sec_otzffi}

For each CCD frame of a science image, we first subtract the associated master bias frame. Then we prepare the science frame for PC-fitting by creating a background map using SExtractor and identify sources in the image for masking from the fit. After subtracting the background, we fit unmasked pixels with a weighted least-squares algorithm, using the PCs associated with the master bias for that night. Again, each half of the CCD is treated separately. The number of PCs included in the fitting procedure is determined by the fraction of the pixels in the row that remain unmasked: all 10 PCs for at least 25\% unmasked; five PCs if 10-25\% of the row is unmasked; one PC if 2-10\% of the row is unmasked; and no fitting is performed if less than 2\% of the row is unmasked.

\begin{figure}
\begin{center}
\includegraphics[width=0.9\columnwidth]{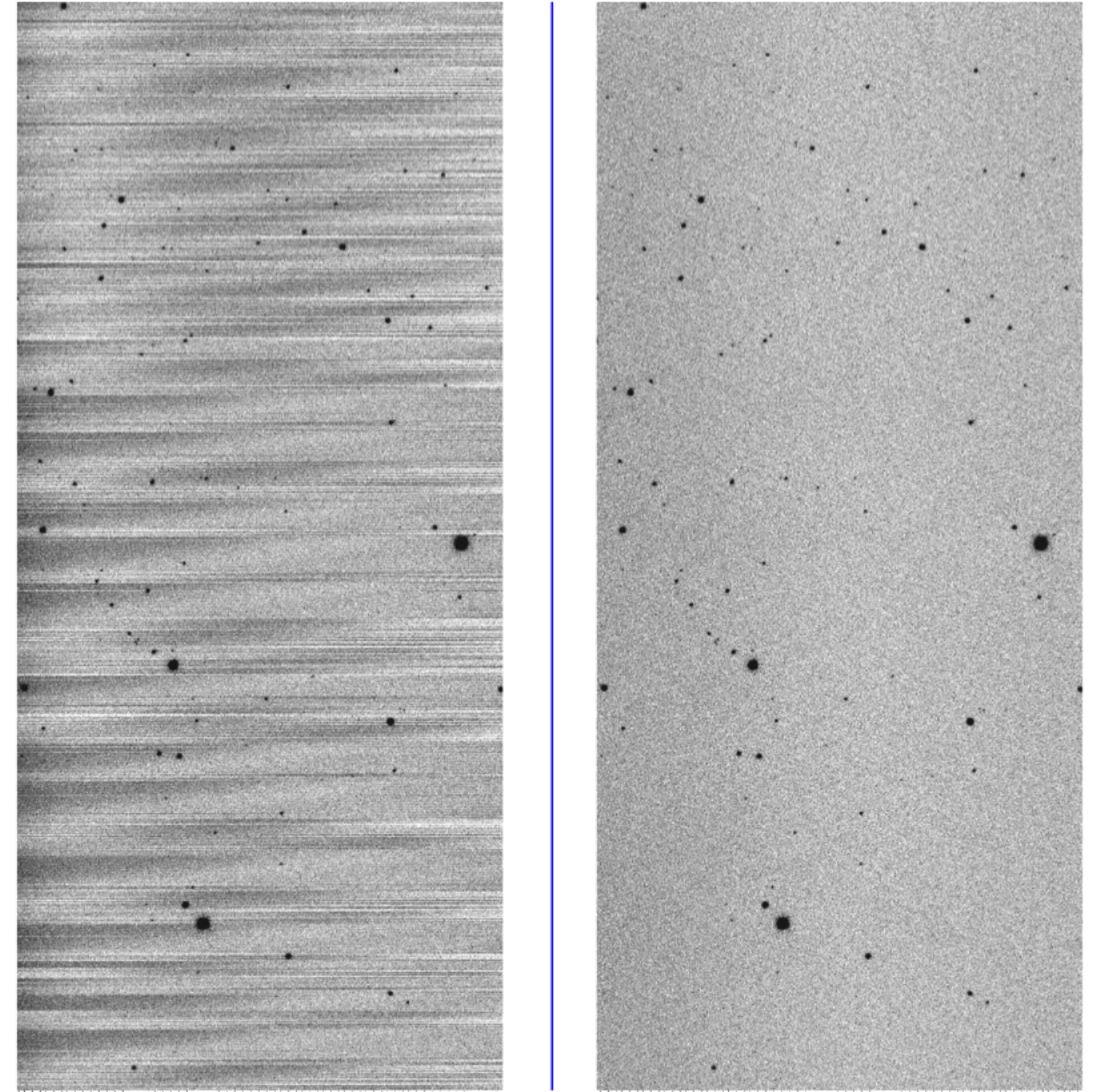}
\caption{Example of bias residuals after traditional overscan correction used in the SkyMapper Early Data Release (EDR, left) vs. PCA model-based bias removal used in DR1 (right). }\label{biasEDR_DR1}
\end{center}
\end{figure}

After the bias PCA treatment, the master flatfield is divided out.

Next, cosmic rays are detected and removed from the science frames using the {\it lacosmicx} software package\footnote{https://github.com/cmccully/lacosmicx}, a fast Python implementation of the L.A.Cosmic routine \citep{2001PASP..113.1420V}. Each amplifier is treated separately to allow for the use of specific read noise settings. From over 7,000 bias frames taken during the DR1 period, we determine the read noise in each frame, and determine for each amplifier a mean level and an rms variation over time. To avoid removal of spurious cosmic rays from uncorrected interference noise in higher-noise frames, we use a high fiducial value for the read noise, setting it to $\langle RON \rangle + 3\sigma_{\rm RON} + 1$ counts for each amplifier. The range of input read noise values is $7.36-17.04$, with a median of 9.95 counts. The saturation level is set to 60,000 counts, with a Laplacian-to-noise limit of 6$\sigma$ and a fractional detection limit for neighbouring pixels of 1.35$\sigma$. The image mask is provided as an input to {\it lacosmicx} in order to avoid detecting known bad pixels. 

We then still observe residual differences in the sky level between the two halves of each CCD due to non-linearity at low count rates. We call SExtractor again to create a background map for each half of the image. From the two background maps, the median value for seven columns near the centre of the CCD (omitting the three columns closest to the middle) are compared, and the difference is then applied as an additive offset to all pixels in the lower of the two halves.

Next, an existing WCS solution is inserted into the image header. The pixel values are then converted to 16-bit integers, with a BZERO value of 32668 and a BSCALE value of 1, to give a resulting count range of -100 to 65435 (preserving the noise near zero counts), and the images and masks are compressed using the lossless {\sc fpack} FITS Image Compression Utility \citep{2009PASP..121..414P} from the CFITSIO library.\footnote{https://heasarc.gsfc.nasa.gov/fitsio/fitsio.html}

Any images for which the raw image could not generate a WCS solution are tried again after the calibrations have been applied. If still no WCS could be obtained, the CCD is discarded for further analysis.

\subsection{Photometry}\label{sec_phot}

For any mosaic image in which at least one CCD has yielded a valid WCS solution, the suitable CCDs are then run through our photometry stage. 

First, SExtractor is run on each CCD. We use a detection threshold of 1.5$\sigma$, the default (pyramidal) filtering function, a minimum area of 5 pixels, a minimum deblending contrast of 0.01, employ a local background estimate, utilise cleaning around bright objects, and measure photometry in a series of apertures (diameters of 4, 6, 8, 10, 12, 16, 20, 30, 40, and 60 pixels). We provide SExtractor with the image-specific mask as a Flag image and a version of the global bad pixel mask as a Weight image. The parameters measured by SExtractor are shown in Table~\ref{tab_sextractor} of the Appendix.

\subsubsection{Aperture Corrections}\label{apercorr}

We adopt the 30-pixel-diameter ($\approx15\arcsec$) photometry as the reference aperture for our aperture corrections. For each of the apertures smaller than 15$\arcsec$, we select candidate sources (those with positive fluxes, no flagged pixels from the mask, SExtractor flags less than 4, a semi-minor axis length of at least one pixel), where we progressively relax constraints on the flux errors and the CLASS\_STAR values if fewer than eight stars on the CCD meet the criteria. Ultimately, if at least five stars are available, the ratio of fluxes between the aperture in question and the reference aperture are fit with a 2-D linear plane, which was found to sufficiently account for any spatial variation in the aperture corrections across a single CCD. The fit is revised through four passes of 2.5$\sigma$-clipping, and the final correction parameters are then applied to all stars on the CCD to yield corrected aperture magnitudes (i.e., predicted 15$\arcsec$-aperture magnitudes based on the smaller apertures). The errors on the corrected fluxes are calculated by square-adding the original flux errors with the RMS around the final planar fit. In cases when fewer than six stars remain unclipped, no gradient is fit and the median aperture correction is adopted for the CCD.

The predicted 15$\arcsec$-aperture magnitudes from the seven smallest apertures are then used to estimate a "PSF" magnitude and error, taken as the weighted mean of the inputs and the error on the weighted mean. The $\chi^2$ value from the fit is also estimated as a way of distinguishing between point sources and other spatial profiles (cosmic rays or extended objects). Note, that these "PSF" magnitudes will be affected by nearby neighbours that contribute flux to the 15$\arcsec$-aperture. 

\subsubsection{Zero-point Calibration}\label{ZPcalib}

For SkyMapper DR1, the zero-point (ZP) calibration is referenced to the AAVSO Photometric All Sky Survey \cite[APASS DR9;][]{2016yCat.2336....0H} and 2MASS. From APASS DR9 we select suitable calibrator stars with clean flags\footnote{For APASS DR9, we require f\_gmag $=0$ and f\_rmag $=0$. For 2MASS, we use: ph\_qual $=$ 'AAA' and cc\_flg $=$ '000' and gal\_contam $=0$ and dup\_src $=0$ and use\_src $=1$.}, magnitude errors below 0.15 and magnitudes in the ranges of $g=[11.5,16.5]$ and $r=[10.5,16]$. From 2MASS we select stars with clean flags, no neighbours within $10\arcsec$ and $K<13.5$. This leaves approximately 12.8 million APASS DR9 sources and 2.5 million 2MASS sources at $\delta<+30\deg$. Then we compute predicted magnitudes in all SkyMapper filters with a method trained on photometry from the first data release of the Pan-STARRS1 3$\pi$ Steradian Survey \cite[PS1;][]{2016arXiv161205560C, 2016arXiv161205243F, 2016arXiv161205242M}.

We first dereddened stars in all three catalogues using the reddening maps by Schlegel, Finkbeiner, \& Davis (1998; hereafter, SFD) and bandpass coefficients derived from a \citet{F99} extinction law. Then we used a large sample of stars that are well-measured in all three catalogues to fit linear relations predicting PS1 magnitude from APASS and 2MASS. As a final step we apply a linear PS1-to-SkyMapper transformation that was derived using the filter curves of PS1 and SkyMapper together with the unreddened F- and G-type main-sequence and sub-giant stellar templates from \citet{1998yCat..61100863P}.

From PS1, we find that trends of mean stellar colour with reddening are minimised if we adopt for the mean reddening of the average star, a prescription of

\begin{equation*}
   \frac{dE(B-V)_{\rm eff}}{dE(B-V)_{\rm SFD}} = 
    \begin{cases}
      0.82, & \text{if}\ E(B-V)_{\rm SFD}<0.1 \\
      0.65, & \text{otherwise}
    \end{cases}
\end{equation*}

\citet{SF11} had already found that the SFD maps overestimate reddening on average and suggest to generally scale them by a factor 0.86. We also find that we need to reduce more strongly the reddening at $E(B-V)_{\rm SFD}>0.1$, and fit a coefficient in order to remove trends of the mean star colour with reddening. We find the same coefficient 0.65 as previously suggested by \citet{2000AJ....120.2065B} for stars of $V=[11,15]$. This scaling probably accounts for the fact that a fraction of the stars reside in front of some of the dust that contributes to the IR emission in the SFD maps. 

We estimate reddening coefficients for our filters using a \citet{F99} extinction law with $R_V=3.1$. This appears to best match the empirical bandpass coefficients found by analysing trends in the observed colours of stars \citep{SF11} and QSOs \citep{W14} with reddening. Our mean bandpass reddening coefficients $R_{\rm band}$ are listed in Tab.~\ref{filtab}, and while they apply strictly only to flat spectra, as $R_{\rm band}$ depends on the object SED, this effect is mostly relevant in the $u$-band.

We restrict calibrator stars to certain colour ranges after de-reddening in order to ensure sufficient linearity of the colour transformation. For calibrating $iz$ images, we use stars with $(J-K)_0=[0.4,0.75]$ and $(g-r)_0=[0.2,0.8]$; for $gr$ images, we require only $(g-r)_0=[0.2,0.8]$ and for $uv$ images, we use $(g-r)_0=[0.4,0.8]$. 

Final SkyMapper AB magnitudes are estimated from APASS AB and 2MASS Vega magnitudes using:

\begin{align*}
  u_2 & =  g_{\rm AP} + 0.622 + 1.757(g-r)_0 + 0.953 E(B-V)_{\rm eff}; \\
  u_1 & =  g_{\rm AP} + 0.575 + 2.050(g-r)_0 + 1.002 E(B-V)_{\rm eff}; \\
  v   & =  g_{\rm AP} + 0.033 + 2.240(g-r)_0 + 0.734 E(B-V)_{\rm eff}; \\
  g   & =  g_{\rm AP} - 0.010 - 0.295(g-r)_0 - 0.306 E(B-V)_{\rm eff}; \\
  r   & =  r_{\rm AP} - 0.003 + 0.042(g-r)_0 + 0.014 E(B-V)_{\rm eff}; \\
  i   & =  J_{\rm 2M} + 0.270 - 0.060(g-r)_0 + 0.060 E(B-V)_{\rm eff} \\
      &      + 0.31(r-J)_0 + 0.207(g-J)_0 + 0.062(J-K)_0; \\
  z   & =  J_{\rm 2M} + 0.550 + 0.062(g-r)_0 + 0.510 E(B-V)_{\rm eff} \\
      &      + 0.880(J-K)_0. \\
\end{align*}

Because of the red leak in the SkyMapper $u$-band, atmospheric extinction makes the filter transformations airmass-dependent (see Fig.~\ref{filtercurves}). We calculate  predicted $u$-band magnitudes for airmasses of 1 and 2, and interpolate between those based on the airmass of each exposure. 

For each SkyMapper source having a positive 15$\arcsec$-aperture flux, a 15$\arcsec$-aperture magnitude error less than 0.04~mag, SExtractor FLAGS < 4 (i.e., not saturated), NIMAFLAGS = 0 (i.e., no masked pixels within the isophotal area), and semi-minor axis size of at least 1 pixel, we find the closest match on the sky from the sample of calibrator stars, with a maximum positional difference of 10$\arcsec$. 

We compute the differences in observed and predicted magnitudes for each star on all available CCDs for the mosaic image, along with the whole-of-mosaic ($x, y$) positions for those stars (adopting the median CCD-to-CCD offsets from $\sim$10,000 images with clean WCS solutions). With the positions and ZP estimates from the entire mosaic, we then simultaneously fit for linear gradients in $x$ and $y$, in order to take atmospheric extinction gradients across the large field-of-view into account.

We perform four passes of 2.5$\sigma$-clipping to determine the final ZP gradient, and fix the gradients to be zero when five or fewer stars remain unclipped. The resulting ZP plane is then applied to the magnitudes for all sources detected in the mosaic. The RMS around the final ZP fit is recorded, but not applied to the individual magnitude errors. Most images in DR1 have between 300 and 1000 calibrator stars per frame that are used for the zeropoint calibration, and in all bands at least 95\% of frames have more then 100 calibrator stars. 

We adjust all $u$-band magnitudes globally 0.05~mag brighter, which makes our measurements consistent with SkyMapper $u$-band predictions derived from measured PanSTARRS $g$- and $r$-bands (see Sect.~\ref{photocomp} for details).

Prior to uploading the photometry into the database, we perform a small amount of cleaning. Because SExtractor does not provide error estimates for the PA, we adopt the PA error used for the \emph{Chandra} Source Catalog: 

\begin{equation*}
	e_{\rm PA} = {\rm atan2}((b+e_b),(a-e_a)) - {\rm atan2}((b-e_b),(a+e_a)) ~, 
\end{equation*}

where $a$ and $b$ are the semi-major axis and semi-minor axis lengths, respectively, and $e_a$ and e$_b$ are their errors;\footnote{See http://cxc.harvard.edu/csc/why/pos\_angle\_err.html} and we impose a minimum PA error of 0.01~deg. We also impose magnitude error floors of 0.0033~mag, and we remove sources having negative fluxes in their 30$\arcsec$ apertures.

\subsubsection{Catalogue Distill}\label{sec_distill}

In a final step we distill the {\it photometry} table with one row per individual detection into a {\it master} table with one entry per unique astrophysical object, using only detections that are not affected by bad flags (FLAGS $<8$). Before the matching, we set the flag with bit value 512 for all detections where any of MAG\_APC05, MAG\_APR15 or MAG\_PETRO is fainter than 19~mag. We also mark unusually concentrated sources such as possible cosmic rays by setting the flag with bit value 1024, and identify these using MAG\_APC02$-$MAG\_APR15 $<-1$ AND CHI2\_PSF $>10$. We mark sources in the vicinity of bright stars by setting the flag with bit value 2048 as these are either spurious or their photometry is affected by scattered light. We consider the environments of stars with $V<6$ in the Yale Bright Star Catalogue \citep{1964cbs..book.....H, 1995yCat.5050....0H} and of stars with $g_{\rm phot}<8$ from Gaia DR1, marking all sources within radii of $10^{-0.2 V}$ or $10^{-0.2 g_{\rm phot}}$ degrees, respectively.

\begin{figure*}
\begin{center}
\includegraphics[angle=270,width=\textwidth]{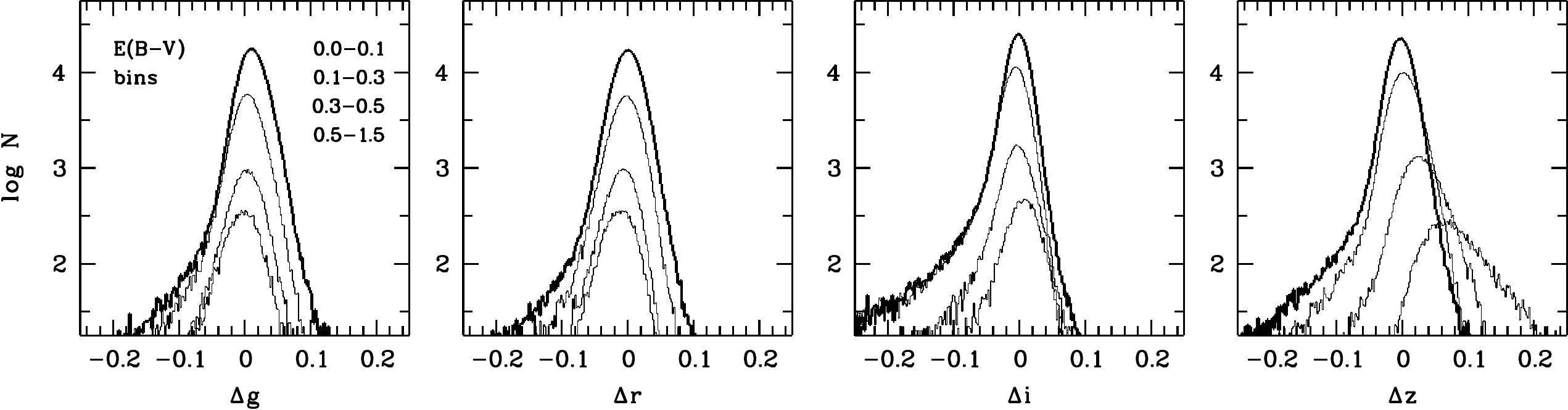}
\includegraphics[width=\textwidth]{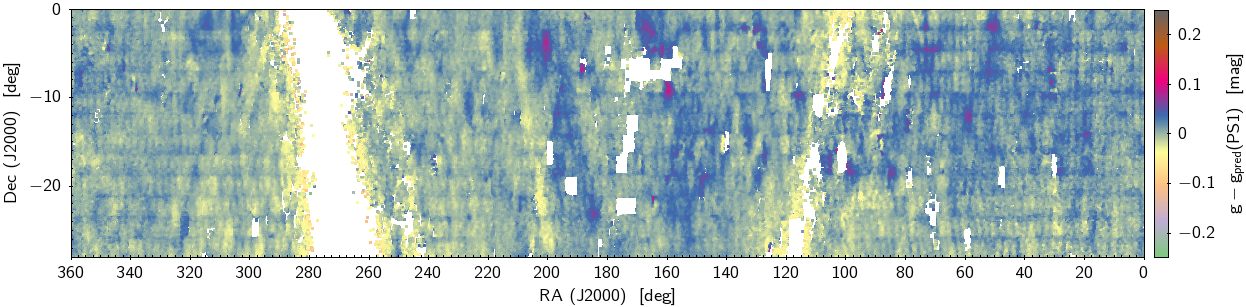}
\includegraphics[width=\textwidth]{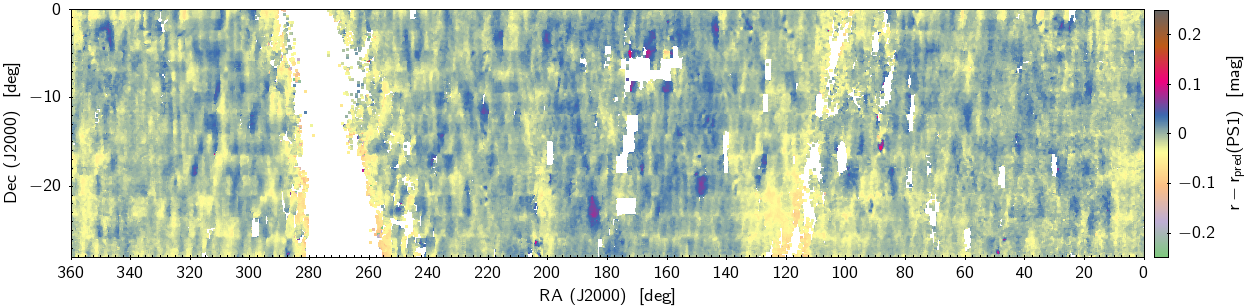}
\includegraphics[width=\textwidth]{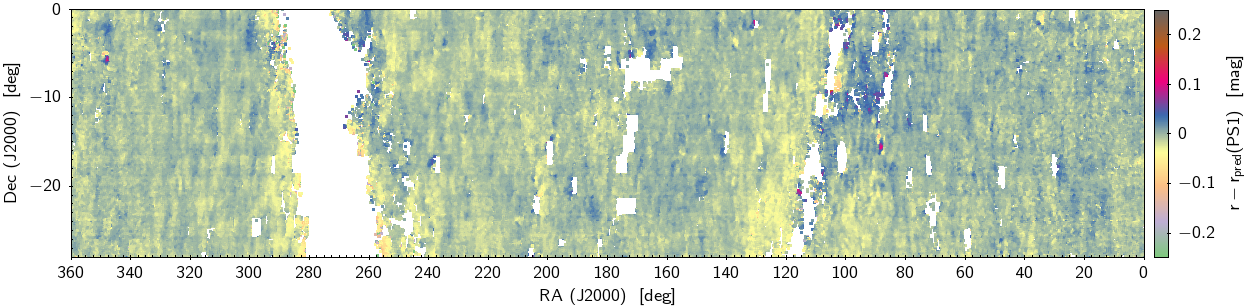}
\includegraphics[width=\textwidth]{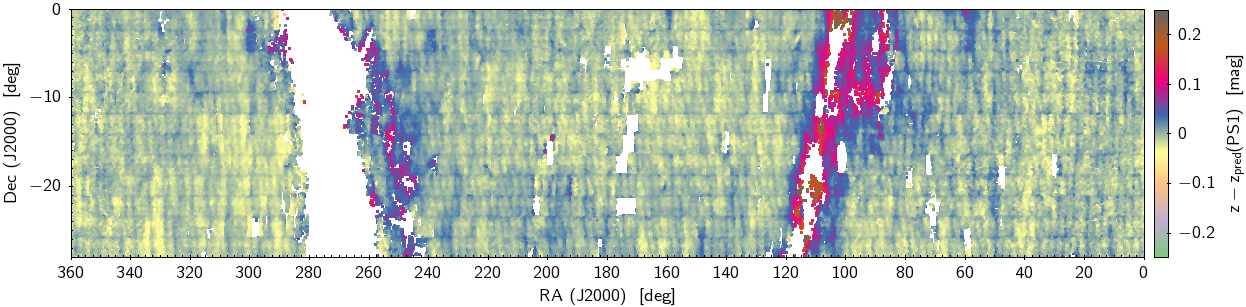}
\caption{Difference between SkyMapper magnitude as measured in DR1 and as predicted from PS1 magnitudes with bandpass transformations applied. This comparison is restricted to calibrator stars that are well measured and have reliable transformations. Top: histograms in bins of interstellar foreground reddening up to $E(B-V)=1.5$. The bottom four panels show spatially resolved sky maps. }\label{SM_PS1_histo}
\end{center}
\end{figure*}

Within each passband, we then position-match detections across different images into groups. We first classify every detection as either a primary or secondary detection, using the latter label when another detection with a higher central count level (FLUX\_MAX) is found on any image within a matching radius of $2\arcsec$. We then associate every secondary detection with the nearest primary detection, which can be over $2\arcsec$ away to accommodate chains of detections. Within each group we combine positions and shapes as unweighted averages, but combine PSF and PETRO photometry as inverse-variance weighted averages. FLAGS are BIT\_OR-combined, NIMAFLAGS are summed up, and from CLASS\_STAR and FLUX\_MAX we pick the largest value. 

Next we merge the per-filter averaged detections between the six filters with a similar procedure. At this stage, however, positions are averaged by weighting them with inverse variance, while shapes, flags and other columns are combined as with the per-filter merging. Most of the time, this process finds only one source per filter (or none) to combine into a unique astrophysical object. Sometimes, however, one astrophysical parent source is associated with two or more child sources in some filter, e.g., a single bright red child source might lie in between two separated faint blue child sources, and be close enough to all be merged into one single parent. In these cases, we attempt to combine the fluxes of the child sources into a total flux, although at this point we do not trust the resulting photometry. Certainly, the PSF photometry can end up counting flux multiple times and thus overestimate the total brightness of a source with multiple children. We always note the number of children per band for each object in the master catalogue, so that it is easy to avoid those sources when selecting complete, photometrically clean samples.

The final master table is cross-matched with several external imaging catalogues, including APASS DR9, UCAC4, 2MASS, AllWISE \citep{WISE, 2011ApJ...731...53M}, PS1 DR1, Gaia DR1, and the GALEX All-sky Imaging Survey \cite[BCScat;][]{2014AdSpR..53..900B}, such that every entry in the SkyMapper table contains an ID pointer to the corresponding object in the external table. The master table is also cross-matched against itself to identify the ID and distance to the nearest neighbour of every source. The maximum matching radius for all cross-matches is $15\arcsec$. In contrast, external spectroscopic catalogues are also cross-matched to the final master table, but such that the external table contains a pointer to the SkyMapper ID, because of the relatively small row numbers in the spectroscopic tables. The cross-matched spectroscopic catalogues include the 6dF Galaxy Survey \cite[6dFGS;][]{2004MNRAS.355..747J, 2009MNRAS.399..683J}, the 2dF Galaxy Redshift Survey \cite[2dFGRS;][]{2001MNRAS.328.1039C}, the 2dF QSO Redshift Survey \cite[2qz6qz;][]{2004MNRAS.349.1397C}, and the Hamburg/ESO Survey for Bright QSOs \cite[HES QSOs;][]{2000A&A...358...77W}. When using cross-match IDs, care needs to be taken to observe the distance column in order to pick only likely physically associated detections. Credible distance thresholds depend on the resolution of the external dataset, e.g., a larger value for AllWISE than for 2MASS, as well as on the epoch difference and proper motion in the case of nearby stars.

\begin{figure}
\begin{center}
\includegraphics[angle=270,width=\columnwidth]{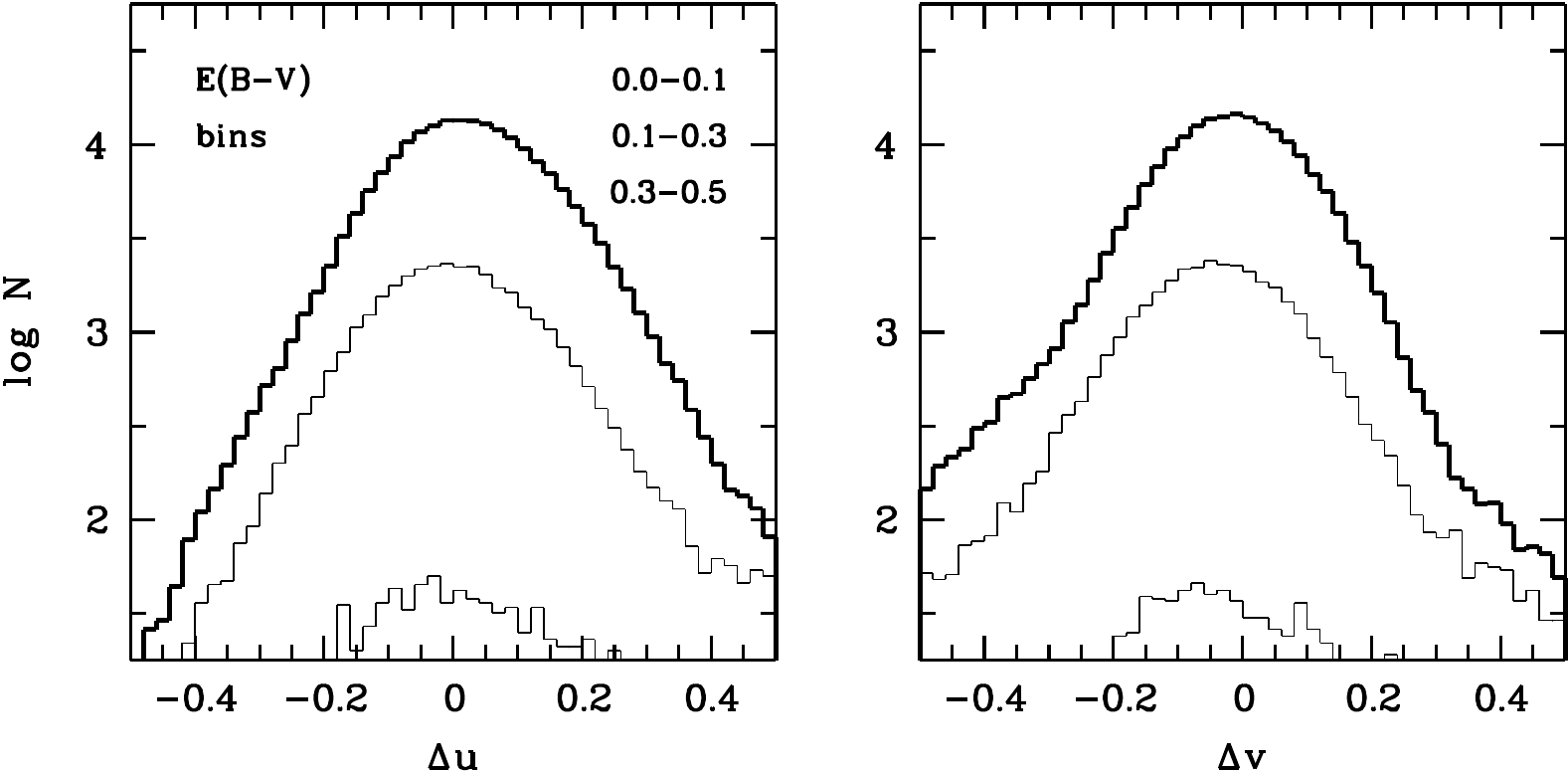}
\caption{Difference between SkyMapper magnitudes and those predicted from PS1 photometry, as in Fig.~\ref{SM_PS1_histo}, but for the highly extrapolated filters $u$ and $v$. The $u$-band has been globally adjusted by $-0.05$~mag, and the $v$-band shows a shift between regions of low vs. high reddening, which likely reflects the metallicity difference between the disk and halo populations in the Milky Way. }\label{sm_ps1_uv}
\end{center}
\end{figure}

\section{Survey properties}

\subsection{Photometric comparison: SkyMapper vs. PanSTARRS1 DR1}\label{photocomp}

First, we assess the internal reproducibility of measured SkyMapper 'PSF' magnitudes of point sources from repeat visits and find that it ranges from 8~mmag rms in $griz$-bands to 12~mmag in the $uv$-bands. For a comparison against external data, we chose to use PanSTARRS1 DR1 and estimate magnitudes in SkyMapper bands from existing measurements in PS1 bands. After checking colour trends with magnitude, we restrict this comparison to calibrator stars that appear well-measured in both surveys, thus ensuring that the transformation from PS1 to SkyMapper bands is reliable\footnote{$g_{\rm SM}=[13.7,15.5]$ and $i_{\rm SM}=[13.6,15.3]$ and $\delta>-28^\circ$}. We deredden the PS1 SED using the modified mean reddening $E(B-V)_{\rm eff}$ specified in Sect.~\ref{ZPcalib}, and apply colour terms fitted on synthetic PS1 and SkyMapper photometry of \citet{1998yCat..61100863P} stars with luminosity class IV and V and colour range $(g-r)_0=[0.2,1.0]$. We apply the following equations to transform measured PS1 $griz$ magnitudes into SkyMapper magnitudes, using a de-reddened PS1 colour of $(g-i)_0=g-i-1.5E(B-V)_{\rm eff}$:

\begin{align*}
  u_2 & = g_{\rm PS} + 0.728 + 1.319(g-i)_0 + 1.075 E(B-V)_{\rm eff}; \\
  u_1 & = g_{\rm PS} + 0.722 + 1.461(g-i)_0 + 1.124 E(B-V)_{\rm eff}; \\
  v   & = g_{\rm PS} + 0.280 + 1.600(g-i)_0 + 0.856 E(B-V)_{\rm eff}; \\
  g   & = g_{\rm PS} - 0.011 - 0.162(g-i)_0 - 0.184 E(B-V)_{\rm eff}; \\
  r   & = r_{\rm PS} + 0.000 + 0.021(g-i)_0 + 0.028 E(B-V)_{\rm eff}; \\
  i   & = i_{\rm PS} + 0.013 - 0.045(g-i)_0 - 0.082 E(B-V)_{\rm eff}; \\
  z   & = z_{\rm PS} + 0.022 - 0.054(g-i)_0 - 0.114 E(B-V)_{\rm eff}. \\
\end{align*}

The $g$-band filter shows the most significant colour term, while the $u$- and $v$-band must be extrapolated from $g$, which is inherently less accurate. The $v$-band magnitude in particular also depends on metallicity and shows clear trends between the halo and disk population. Indeed, the motivation for the SkyMapper $v$-band is to act as a metallicity and surface gravity indicator for stars, and below we demonstrate the first step in our search for extremely metal-poor stars (see Sect.~\ref{EMP_sel}). 

\begin{figure}
\begin{center}
\includegraphics[angle=0,width=\columnwidth]{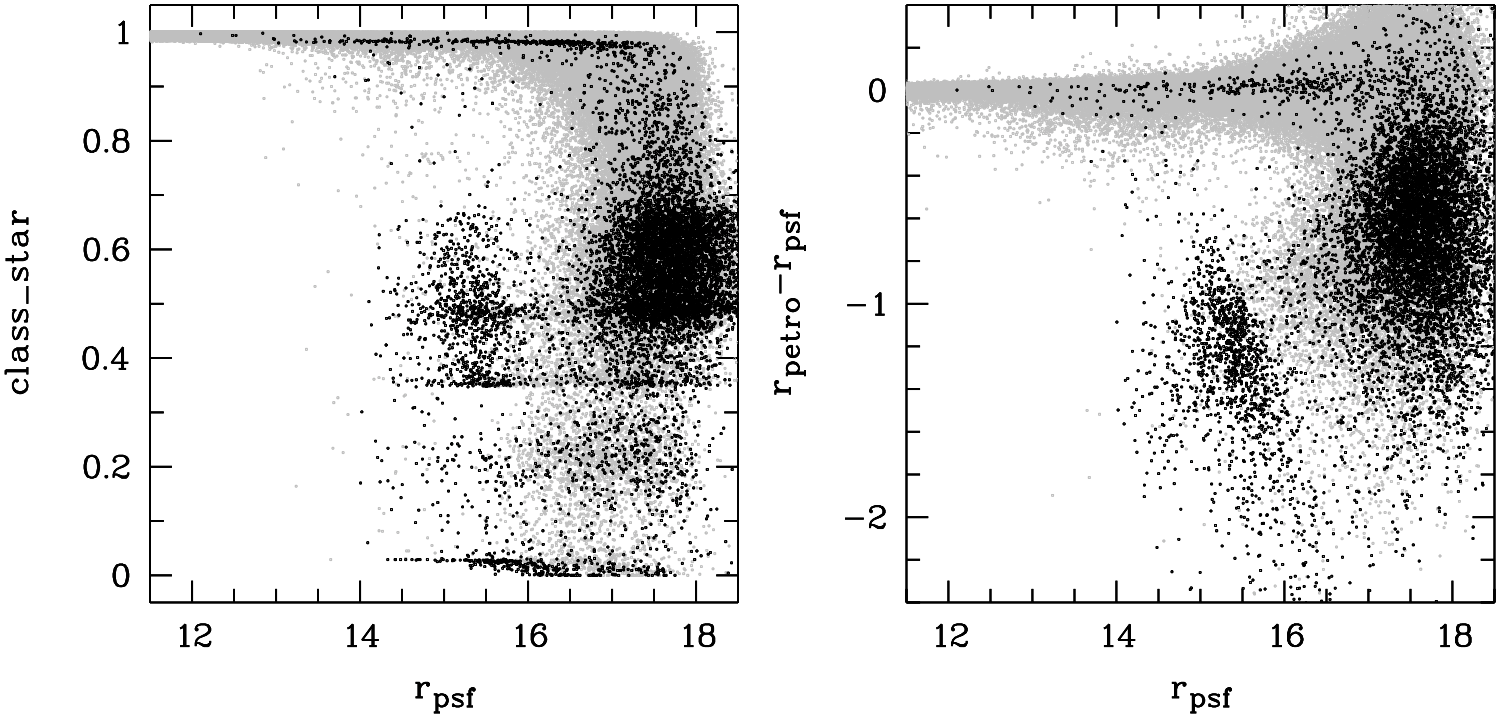}
\caption{Morphology indicators vs. $r$-band magnitude: {\it Left:} SExtractor stellarity index CLASS\_STAR. {\it Right:} Difference between PSF and Petrosian magnitude. Plotted are mag-limited galaxy samples from 6dFGS and 2dFGRS (black) and matching lower-density samples of random DR1 objects (grey). }\label{SM_morph}
\end{center}
\end{figure}

In Fig.~\ref{SM_PS1_histo} we show the difference between the measured and PS1-predicted SkyMapper photometry in histograms after binning in foreground reddening, as well as in sky maps. We restrict this comparison to the four non-extrapolated filters $griz$, and find an rms scatter in the magnitude difference of 23~mmag. The mean offset is generally less than 10~mmag, with the exception of a tail at higher reddening in $z$-band, where the median difference increases to 0.06~mag for $E(B-V)=[0.5,1.5]$. The sky maps also show the bias in $z$-band towards low Galactic latitude where reddening tends to be higher. Apart from that the sky maps show low-level discontinuities at the SkyMapper field edges. SkyMapper fields are arranged in declination stripes with small overlaps in RA and Dec. However, additional passes of the Shallow Survey after the DR1 imaging period provide field overlaps to support a more homogeneous calibration.

Despite the uncertainties introduced by extrapolating the SkyMapper $u$- and $v$-band, we can formally compare the photometry. Again we derive colour terms for SkyMapper $u$ at airmass 1 and 2, although in this comparison we use the median airmass of 1.11 to represent the whole comparison sample. We show the histograms of magnitude differences in Fig.~\ref{sm_ps1_uv} and note that the rms scatter is 0.12~mag in both bands. After finding a mean offset of 0.05~mag in $u$ we adjusted all our $u$-band magnitudes globally 0.05~mag brighter. We assume that this offset has originated from our calibration procedure, where we extrapolated the APASS DR9 $gr$ photometry to the SkyMapper $u$-band. Even though we see slight offsets in $v$-band and a $\sim 0.05$~mag difference between regions of reddening less than $E(B-V)=0.1$ and those with more reddening, we do not correct these. The shift with reddening is likely to be a metallicity effect, where halo and disk stars are expected to have different $v-g$ colours by roughly the observed amount. 

\begin{figure}
\begin{center}
\includegraphics[angle=270,width=0.7\columnwidth]{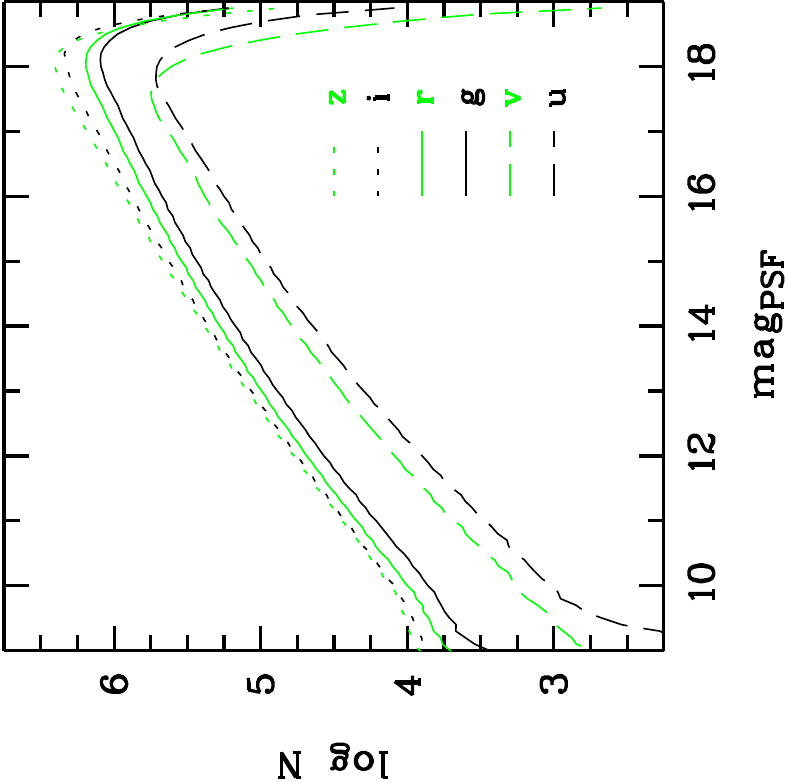}
\caption{Number counts in PSF magnitude for each band. Except for the $v$-band, the completeness limit is nearly 18~mag.}\label{NC6}
\end{center}
\end{figure}

\begin{figure*}
\begin{center}
\includegraphics[width=\textwidth]{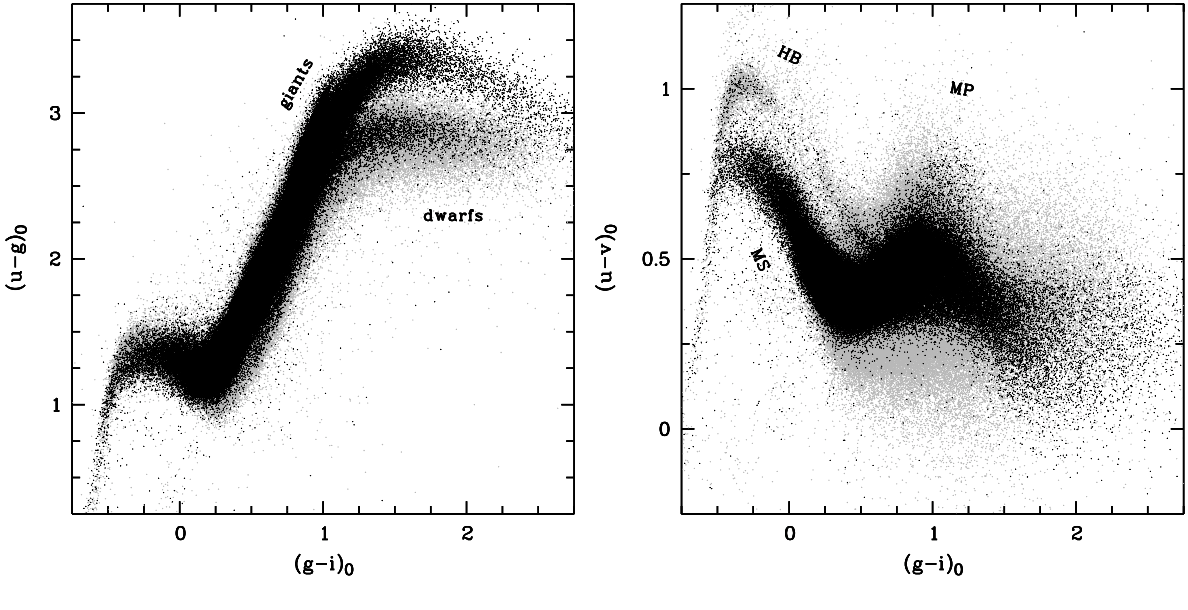}
\includegraphics[width=\textwidth]{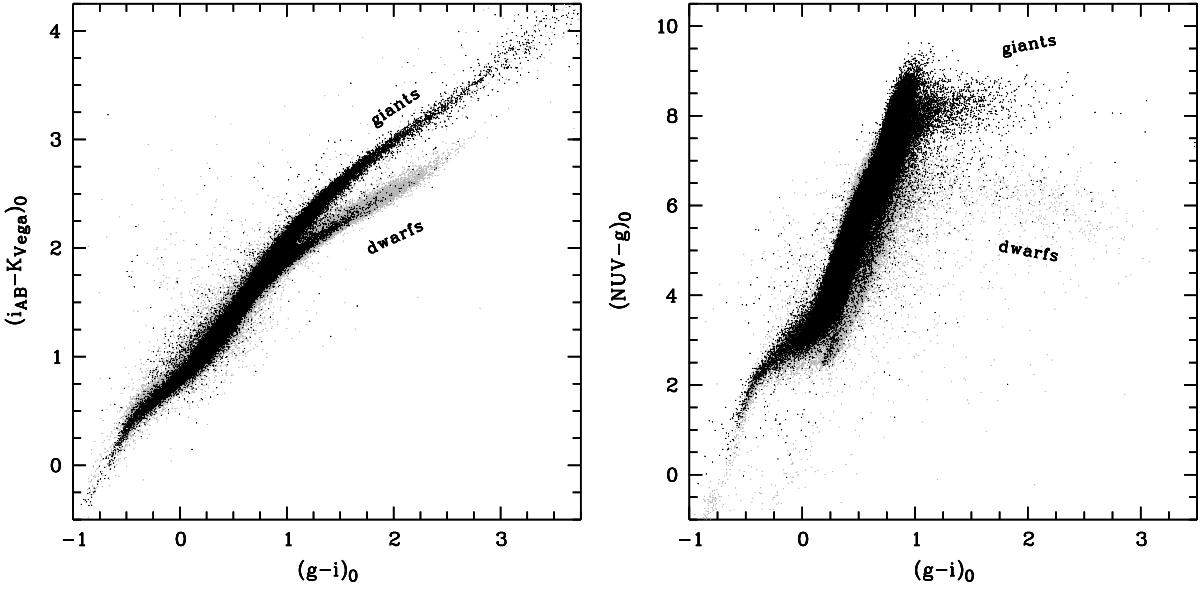}
\caption{De-reddened SkyMapper colours measured for stars with $E(B-V)<0.1$ and $r_{\rm PSF}=[12,12.5]$ (black) or $r_{\rm PSF}=[14,14.5]$ (grey). {\it Top:} Internal SkyMapper colours -- we can easily see the bifurcation between cool dwarfs and cool giants in the top left panel, and separate horizontal branch (HB) stars from main-sequence (MS) stars in the top right panel; also shown is a region where metal-poor (MP) stars can be found easily. {\it Bottom:} Combined colour between a SkyMapper band and a 2MASS band or a GALEX band, respectively. }\label{SM_cc4}
\end{center}
\end{figure*}

\subsection{Morphology indicators}

Separating point sources from extended sources is a challenge limited by the spatial resolution set by observing conditions and by PSF variations within a given frame. Any indicator of morphology will work best with bright sources but be increasingly ambiguous for noisier faint sources. We consider two morphology indicators: (i) the SExtractor stellarity index CLASS\_STAR, which approaches 1.0 for point sources and typically has values between 0 and 0.9 for extended sources, and (ii) the difference between the PSF and Petrosian magnitudes. 

In Fig.~\ref{SM_morph} we show these two morphology indicators vs. the $r$-band magnitude for a large sample of objects including bright galaxies from the hemispheric 6dFGS and faint galaxies from the deeper and smaller-area 2dFGRS (shown together with black points). We restrict the plotted sample to objects with good flags and no nearby neighbours. We note that both galaxy surveys contain compact galaxies, e.g. emission-line galaxies with high star-formation densities and active galaxies with dominant Seyfert-1 nuclei. In grey we show a random sample of all DR1 objects matched in brightness to those two galaxy surveys. The two measures of morphology appear largely similar except for high-stellarity sources at the faint end: SExtractor tends not to attribute a high likelihood of stellarity to faint, noisy sources, while the difference between PSF and Petrosian magnitude for true point sources always scatters around zero, no matter the level of noise. Thus, CLASS\_STAR tends to select proven point sources with low contamination and is incomplete at the faint end, while the magnitude comparison selects possible point sources with high completeness but increasing contamination at the faint end. 

The fs\_photometry table also contains for each separate detection our point-source likelihood measure CHI2\_PSF, which is closer to the magnitude comparison in behaviour. It essentially expresses in $\chi^2$ units, how well the sequence of aperture magnitudes, i.e. growth curve, follows the local model of a PSF growth curve.

\subsection{Survey depth and number counts}

As DR1 includes only images from the Shallow Survey, it is expected to be $\sim4$~mag shallower than the full SkyMapper survey. In Fig.~\ref{NC6} we show number counts per filter from our master table. These typically turn over around 18~mag in each band, with the exception of the shallower, relatively narrow, $v$-band that appears complete to 17.5~mag. Galaxies are included in this plot although they will be incomplete already at much brighter magnitudes given that they are more extended than a PSF and their outer parts will get lost in the noise of the Shallow Survey images.

A given mean population of stars can be roughly characterised by a horizontal line in Fig.~\ref{NC6}. The horizontal offset of the number count lines then shows the mean colours of the stellar population. This suggests e.g. that a $v$-selected population with $v<17.5$ will only be largely complete to $z<15$.

\begin{figure}
\begin{center}
\includegraphics[angle=270,width=\columnwidth]{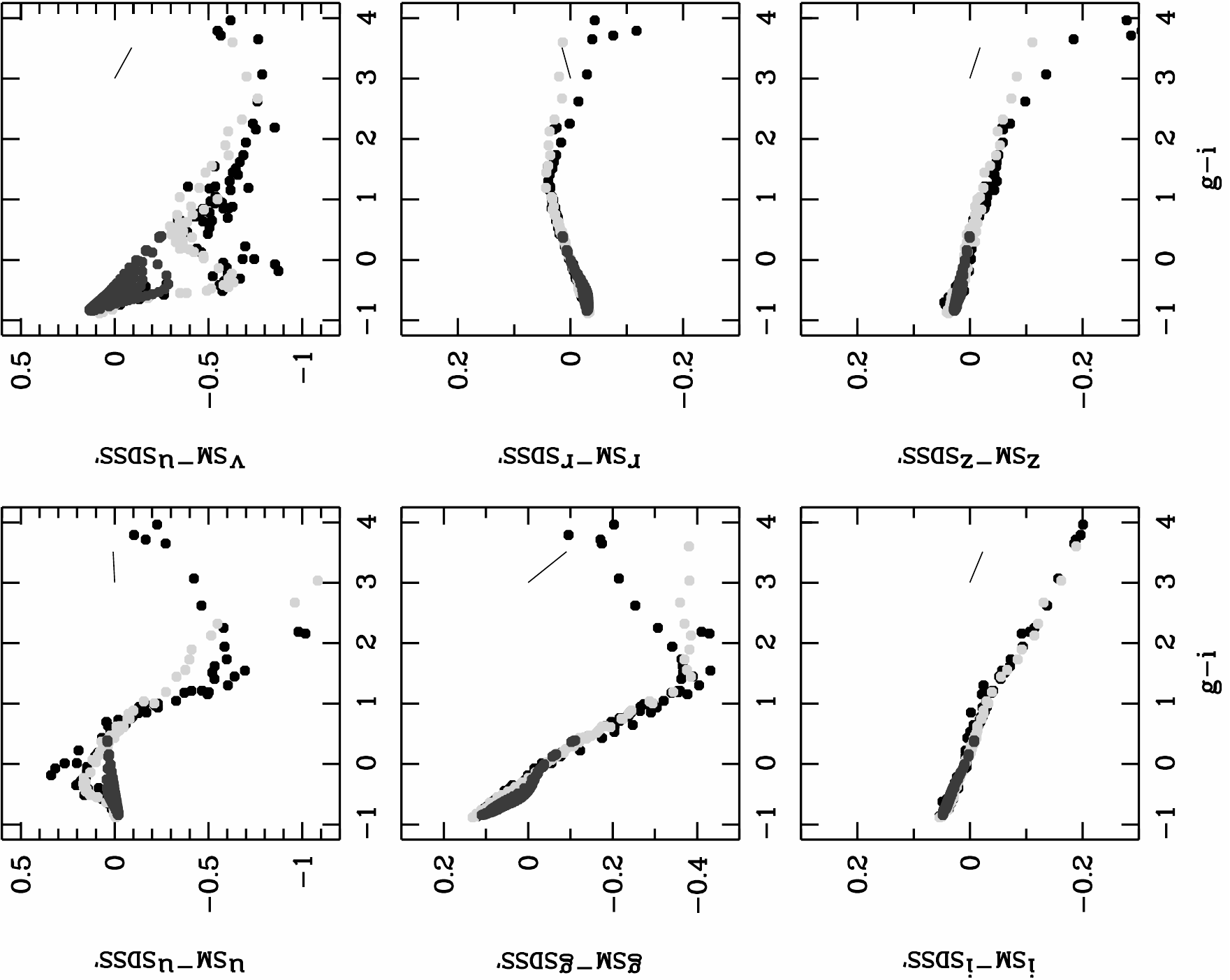}
\caption{Colour terms between SkyMapper and SDSS for the spectral libraries of main-sequence (IV/V, black points) and giant (I-III, light grey points) F-M stars of \citet{1998yCat..61100863P}, and a grid of DA/DB white dwarf spectra from $T_{\rm eff}= [6000, 40000]$~K by Detlef Koester (priv. comm; dark grey points). The line shows a reddening vector for $A_V=1$ using a \citet{F99} law.}\label{SM_colcorrs}
\end{center}
\end{figure}

\subsection{Colour space and colour terms for stars}

In Fig.~\ref{SM_cc4} we explore the locus of point sources in several de-reddened colour-colour diagrams. We select stars from two magnitude bins ($r=[12,12.5]$ and $r=[14,14.5]$) with low reddening $E(B-V)<0.1$. The first two panels demonstrate how SkyMapper colours alone can differentiate between cool dwarfs and cool giants using the $u-g$ colour, and between AF-type main-sequence stars and horizontal-branch (HB) stars using the $u-v$ colour. The SkyMapper-specific $u$ and $v$ bands bracket the Hydrogen Balmer break at 365~nm, and lower surface gravity causes narrower Balmer absorption lines and thus higher $v$-band flux and a redder $u-v$ colour.

The bottom two panels exploit the fact that our database holds copies of the 2MASS and GALEX BCS catalogues \citep{2014AdSpR..53..900B}, to which we have pre-cross-matched all DR1 sources. Hence, multi-catalogue colours can be obtained with simple and fast table joins. In de-reddening GALEX NUV magnitudes we used $R_{\rm NUV}=7.04$, which we derived by adding the empirical $E(NUV-g)$ coefficient determined by \citet{2013MNRAS.430.2188Y} to the $R_g$ of SDSS. The separation of the cool dwarf and giant branches is also clearly visible in optical-NIR colours such as $i-K$. Optical-to-near-UV colours such as $NUV-g$ will further help to differentiate stars from galaxies and QSOs, when morphology or other indicators are unreliable.

Finally, in Fig.~\ref{SM_colcorrs} we show colour terms between the SkyMapper and SDSS bandpasses, from synthetic photometry performed on all Pickles stars (from main-sequence stars to supergiants) as well as on a grid of white dwarf spectra from Detlev Koester (priv. comm.). We compare the SkyMapper $u$- and $v$-filters both to the SDSS $u^{\prime}$-band, and the other four filters to their eponymous ($g^{\prime} r^{\prime} i^{\prime} z^{\prime}$) SDSS cousins. Between those cousins, the $g$-band shows the strongest colour terms, which reach up to 0.4~mag difference for cool stars.

\begin{figure*}
\begin{center}
\includegraphics[width=1.8\columnwidth]{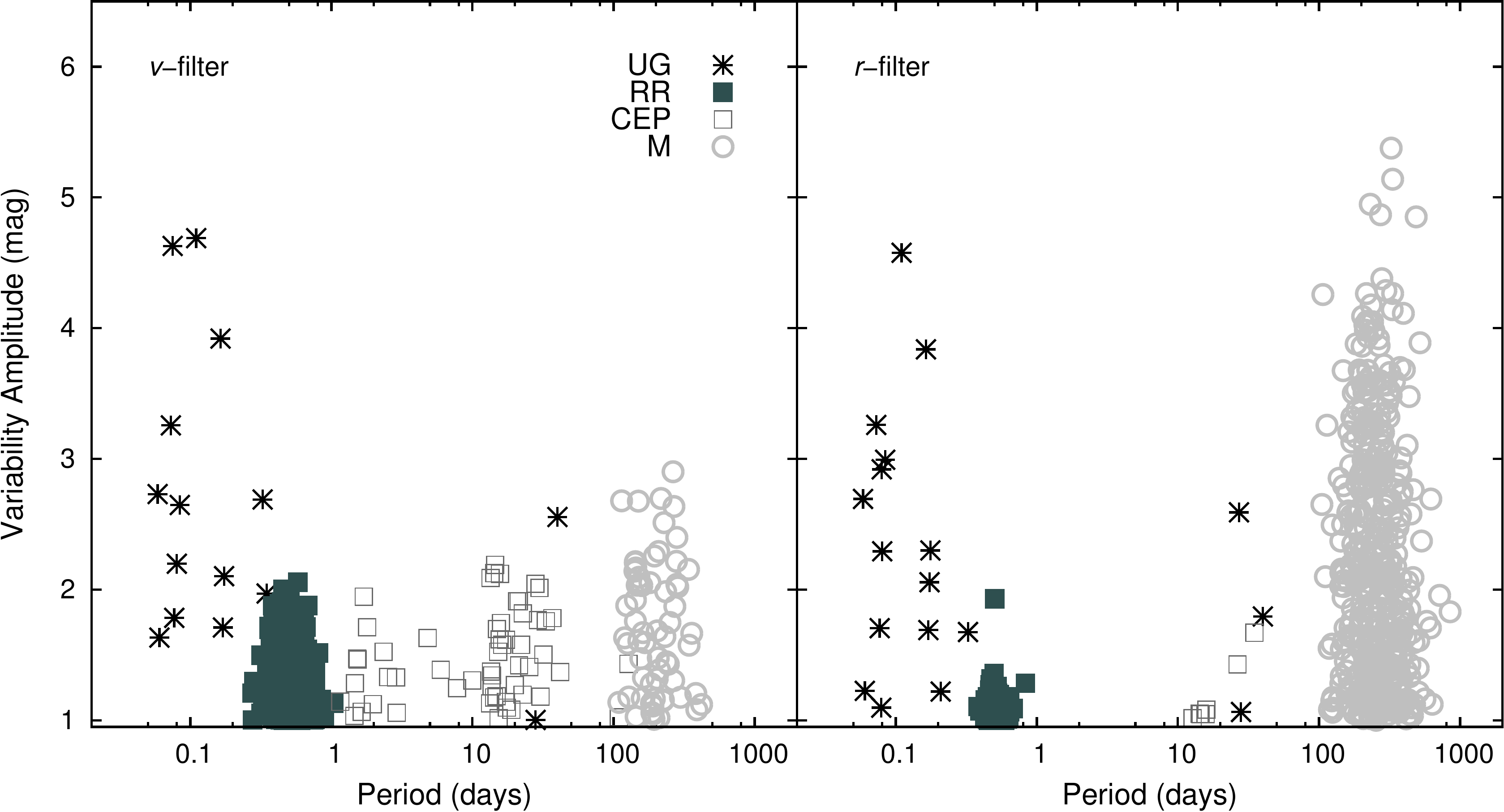}
\caption{Variability amplitude of $\sim$3,000 known variable stars from the AAVSO International Variable Star Index (VSX) with matches in DR1 and amplitudes in either $v$-band or $r$-band of over 1~mag. For clarity this figure is restricted to U Geminorum stars (UG), RR Lyrae stars (RR), Cepheids (CEP) and Mira variables (M). }\label{SM_variables}
\end{center}
\end{figure*}

\section{Data Access and Example Applications}\label{examples}

The DR1 dataset is available to the community through the SkyMapper Node\footnote{http://skymapper.anu.edu.au} of the All-Sky Virtual Observatory\footnote{http://www.asvo.org.au} (ASVO). The sky covered in the release can be browsed at the Node website, either in colour or in each individual SkyMapper passband, with the SkyViewer, a visual exploration tool based on Aladin Lite \citep{AladinLite} and Hierarchical Progressive Survey methods \citep[HiPS;][]{HiPS}. The SkyMapper Node provides further tools that conform to the standards of the International Virtual Observatory Alliance\footnote{http://www.ivoa.net} (IVOA), and includes catalogue access through the Table Access Protocol (TAP) and a Simple Cone Search function, as well as access to the reduced images (and their masks) through the Simple Image Access protocol (SIAP). Note that TAP queries are currently limited to returning one million rows at most.

The TAP tables are arranged in several distinct schemas, which group together the tables of each DR, plus the 'ext' schema of external catalogues. The layout of tables and the details of the SkyMapper table columns are provided in the Appendix. In the following, we discuss a few example applications together with the associated database TAP queries.

\begin{table*}
\caption{Candidates of low-redshift changing-look Seyfert galaxies identified by comparing SkyMapper DR1 $g$-band magnitudes with $B_J$ photometry reported by the Hamburg-ESO Survey \citep{2000A&A...358...77W}. Three out of five objects have changed from type-1 to (nearly) type-2 (quoted 2017 types are provisional).}
\label{Sy12_list} 
\centering          
\begin{tabular}{lcrcccl}
\hline\hline       
HEname & $z$ & DR1 object\_id & $B_J$ (HES) & $g_{\rm petro}$ (DR1) & $\Delta m$ & 2017 type\\ 
\hline
 HE 0311-3517  & 0.114 &  10275697 & 15.59 &  16.497 &  -1.177 & Sy-1.2 \\
 HE 0345-3033  & 0.095 &  10693348 & 16.76 &  18.319 &  -1.829 & Sy-1.9 \\
 HE 1101-0959  & 0.186 &  58455388 & 16.79 &  17.608 &  -1.088 & Sy-1   \\
 HE 1416-1256  & 0.129 &  66593331 & 16.44 &  17.592 &  -1.422 & Sy-1.8 \\
 HE 1514-0606  & 0.133 & 100043473 & 17.07 &  18.120 &  -1.320 & Sy-2   \\
\hline                  
\end{tabular}
\end{table*}

\subsection{Properties of known variable stars}

We investigated 256,146 known variable stars from the AAVSO International Variable Star Index \citep[VSX;][]{VSX06,VSXcat} with cross-matches in DR1 and look for brightness variations seen by SkyMapper repeat visits. We obtain this information, e.g. for the $r$-band, from the database using the query:

\begin{verbatim}
SELECT name,type,period,object_id,
       MAX(f.mag_psf)-MIN(f.mag_psf) AS dm
  FROM ext.vsx 
  JOIN dr1.master
       ON (dr1_id=object_id AND dr1_dist<10)
  JOIN dr1.fs_photometry f 
       USING (object_id)
 WHERE f.flags<4 AND f.NIMAFLAGS=0 AND
       nch_max=1 AND filter='r'
 GROUP BY name,type,period,object_id;
\end{verbatim}

Here, we avoid objects where the photometry may be affected by a range of effects captured in the FLAGS column, by bad pixels as captured in the NIMAFLAGS column, and by a merger of child objects (NCH\_MAX).

In Fig.~\ref{SM_variables} we show variability amplitudes vs. the known period or variability time scale of the stars, for all matches that have over 1~mag difference between the brightest and faintest good-quality detection in DR1. This information is of course available for all six SkyMapper bands, but we show only the $v$- and $r$-band to illustrate how the variability amplitude depends on the type of variable star and its temperature evolution.

The combined high-amplitude list from these two bands contains 39 U Geminorum stars, 2,200 RR Lyrae stars, 55 Cepheids and 520 Mira variables. Here, we omit further objects from rarer types and those with less characteristic time-scales, such as eclipsing binaries. The variables thus end up clearly grouped by their types, which have characteristic and different time scales.

\subsection{Long-term variability of QSOs}

We investigated the variability of QSOs and Seyfert-1 galaxies identified by the Hamburg-ESO Survey \citep[HES;][]{2000A&A...358...77W}. The public HES catalogue lists $B_J$ magnitudes measured on photographic plates exposed over 20 years ago, which we compare to the SkyMapper $g$-band magnitudes observed in 2014/5 to study long-term variability. We can obtain a list of HES QSOs and their approximate magnitude change with a simple query. 

However, with a more specific aim in mind we look for candidates of low-redshift changing-look Seyfert galaxies; given that HES only includes broad-line AGN, this search is only sensitive to transitions from Sy-1 to Sy-2, which would manifest themselves by a large drop in brightness. We first restrict the list to redshifts below 0.3 and find 133 objects with a mean $B_J-g_{\rm petro}$ difference of $+0.27$~mag due to Vega-to-AB and bandpass differences, and an RMS scatter of 0.51~mag.

Taking the mean difference as indicating no variability, we search for objects with a brightness decline of more than 1~mag by selecting for $B_J-g_{\rm petro}-0.27<-1$ using the query:

\begin{verbatim}
SELECT hename,z,object_id,bjmag,g_petro,
       bjmag-g_petro-0.27 AS dm
  FROM ext.spec_hesqso 
  JOIN dr1.master
       ON (dr1_id=object_id AND dr1_dist<3)
 WHERE flags<4 AND NCH_MAX=1
       AND bjmag-g_petro-0.27<-1 AND z<0.3;
\end{verbatim}

This query lists five candidates (see Tab.~\ref{Sy12_list}), which we have all followed up spectroscopically with the ANU 2.3m-telescope and the WiFeS spectrograph \citep{WiFeS} in late 2017. All five Sy1-galaxies show a drop in UV continuum and the brightness of their broad Hydrogen emission lines, when compared to HES spectra taken two decades earlier. Three out of the five show little or no visible broad H$\beta$-emission any more, and thus have changed their type from Sy-1 to Sy-2. 

\begin{figure*}
\begin{center}
\includegraphics[width=\textwidth]{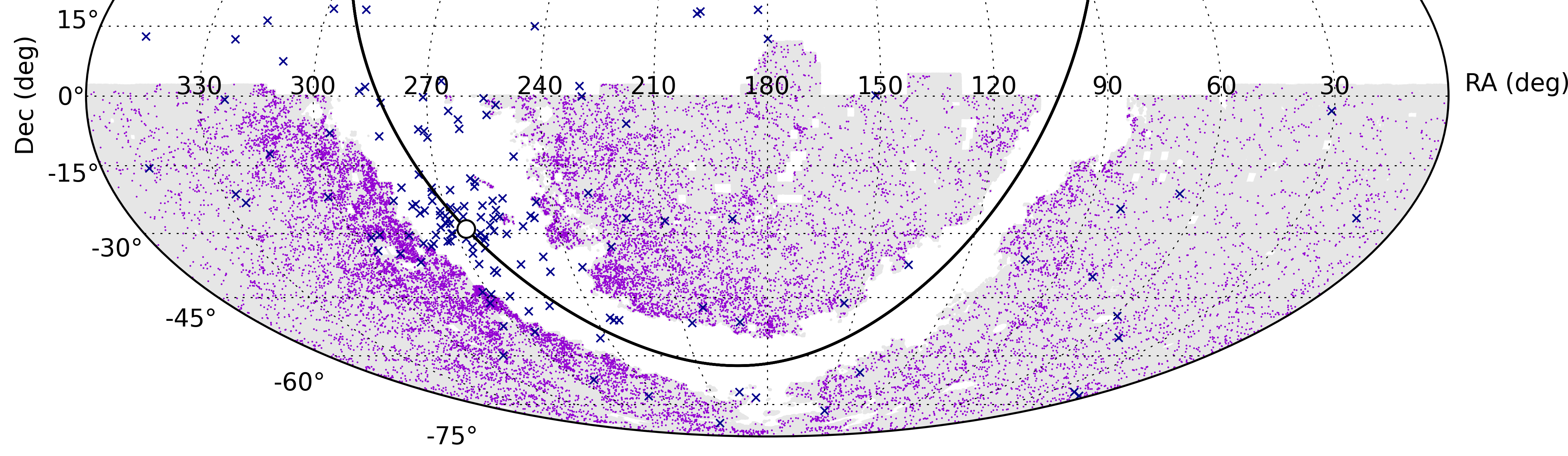}
\caption{Map of EMP candidates with DR1 photometry suggesting $[$Fe$/$H$]<-2.5$: the shaded grey area marks the region from which candidates were selected after applying a selection for stars with high-quality photometry (see Sect.~\ref{EMP_sel}). The plane of the Milky Way is indicated with a solid line, and the Galactic centre with a large open circle. Known globular clusters are marked with blue crosses. }\label{EMPmap}
\end{center}
\end{figure*}

\subsection{Selecting metal-poor star candidates}\label{EMP_sel}

Identifying a large sample of chemically pristine stars in the Milky Way is a prime science driver for SkyMapper and a key motivation behind the distinct $u$- and $v$-filters \citep[e.g.][]{Keller07,B11}. The $v$-filter, in particular, was designed to be metallicity-sensitive, especially at low metallicities where the greatly decreased significance of the Calcium H and K lines translates directly to enhanced flux. The release of SkyMapper DR1 facilitates a uniform search for extremely metal-poor (EMP) stars across the entire Southern hemisphere; the comparatively bright limit of the Shallow Survey is ideally suited to this task as high-resolution spectroscopic follow-up of stars fainter than magnitude $\sim 16$ is prohibitively expensive even on $8-10$m class facilities. 

To identify EMP candidates we utilise a 2-colour plot involving the $g-i$ colour, which serves as a proxy for effective temperature and allows us to isolate stars of the desired spectral type, and the $v-g$ colour, which indicates metallicity. On the plane defined by these two colours, stars with $[$Fe$/$H$]<-2.5$ are separated from the vast bulk of more metal-rich sources by more than $\sim 0.5$~mag; for giants with ${\rm T}_{\rm eff} \approx 5000$K the spread between stars with $[$Fe$/$H$] \approx -2.5$ and $[$Fe$/$H$] \approx -4.0$ is $\sim 0.1$~mag. This selection strategy facilitated the discovery, from SkyMapper commissioning data, of the most iron-deficient star known \citep{Keller14}, which has $[$Fe$/$H$] \leq -6.5$ \citep{Nordlander17}. Our current DR1-based EMP survey is yielding stars at $[$Fe$/$H$] \leq -3.0$ with $\sim 35\%$ efficiency (see Da Costa et al. 2018, in prep.).

Before separating stars according to their colours we select stellar objects with high-quality photometry from an area with reddening limited by $E(B-V)<0.2$. We impose cuts on the $g$-band magnitude ($<16$), the photometric uncertainties in $v$, $g$, and $i$ (0.05, 0.03, and 0.03~mag, respectively), the CLASS\_STAR index ($>0.9$), the flags indicating any potential issues with the measurement (FLAGS $<3$ and NCH\_MAX $=1$), and the location of the nearest neighbour ($>7.5\arcsec$), and we insist on at least one clean detection in all six filters and two in v (given its central importance). This translates into the query:

\begin{verbatim}
SELECT *
  FROM dr1.master 
 WHERE g_psf<16 AND flags<3 AND NCH_MAX=1 AND
       e_v_psf<0.05 AND e_g_psf<0.03 AND
       e_i_psf<0.03 AND class_star>0.9 AND
       prox>7.5 AND ngood_min>=1 AND v_ngood>1
       AND ebmv_sfd<0.2;
\end{verbatim}

This defines a sample of 7.7 million sources, which cannot be downloaded in one step due to the current million-row limit for TAP queries. A subsequent colour analysis thins the list to $\approx 25\,000$ high-quality candidates, those where the $v-g$ colour suggests $[$Fe$/$H$] \leq -2.5$. Their spatial distribution is shown in the all-sky map of Fig.~\ref{EMPmap} and exhibits the concentration towards the Galactic centre expected for a halo-like population, as well as some degree of sub-structure -- although whether this is real or an artefact of residual imperfections in the photometric calibration is not yet understood (for example, note a deficiency in the number of candidates along a clear strip near $\alpha \approx 280$~deg and $\delta \approx -65$~deg).

\section{Survey Future}

Since March 2015, observations with SkyMapper have focused on the deeper exposures for the Main Survey. Further visits of the Shallow Survey have continued during moon phases of $>94$\%. The additional fourth and fifth passes of the Shallow Survey are dithered with much larger offsets than the first three visits; the first three are only intended to fill the gaps in the CCD mosaic, but the further visits involve approximate half-field dithers to assist with future plans to make the calibration more homogeneous. This can be implemented with an \"{u}bercal approach, where measurements for common objects in overlapping parts of images help to constrain a fine-tuning of all image zeropoints \citep[see also][]{ubercal}.

The Main Survey observes full colour sequences that are collected within 20 minutes, as well as additional $gr$ image pairs added during dark time, and $iz$ image pairs primarily added during astronomical twilight. Since the Shallow Survey images are mostly limited by read-out noise, the Main Survey will be a full 4~mag deeper. We plan that, together, they will provide multiple independent and reliable measurements for over 99\% of astronomical objects in the Southern sky at magnitudes ranging from 9 to 22. There will then also be more emphasis on extended-source and forced-position photometry. 

However, the Shallow Survey presented in this paper will eventually serve as the calibration source for the Main Survey; it contains mostly point sources and many variable objects. Hence, this work has focused on instantaneous point-source photometry. The scientific analysis of DR1 has progressed over several topics, and spectroscopic follow-up of many sources has been started. This new dataset is already proving to be a treasure map for astronomers.


\begin{acknowledgements}
The national facility capability for SkyMapper has been funded through ARC LIEF grant LE130100104 from the Australian Research Council, awarded to the University of Sydney, the Australian National University, Swinburne University of Technology, the University of Queensland, the University of Western Australia, the University of Melbourne, Curtin University of Technology, Monash University and the Australian Astronomical Observatory. SkyMapper is owned and operated by The Australian National University's (ANU) Research School of Astronomy and Astrophysics (RSAA). The survey data were processed and provided by the SkyMapper Team at ANU. The SkyMapper node of the All-Sky Virtual Observatory is hosted at the National Computational Infrastructure (NCI). Parts of this project were conducted by the Australian Research Council Centre of Excellence for All-sky Astrophysics (CAASTRO), through project number CE110001020. BPS acknowledges support from the ARC Laureate Fellowship FL0992131. We acknowledge support from the ARC Discovery Projects DP0343962 (BPS, MSB), DP0878137 (BPS, MSB, GDC),  DP120101237 (GDC, DM), and DP150103294 (GDC, DM). DM also acknowledges support from an ARC Future Fellowship FT160100206. This project was undertaken with the assistance of resources and services from the NCI, which is supported by the Australian Government. 

This work has made use of data from the European Space Agency (ESA) mission {\it Gaia}\footnote{https://www.cosmos.esa.int/gaia}, processed by the {\it Gaia} Data Processing and Analysis Consortium\footnote{https://www.cosmos.esa.int/web/gaia/dpac/consortium} (DPAC). Funding for the DPAC has been provided by national institutions, in particular the institutions participating in the {\it Gaia} Multilateral Agreement. This paper makes use of data from the AAVSO Photometric All Sky Survey, whose funding has been provided by the Robert Martin Ayers Sciences Fund. We thank Sara Beck and Arne Henden from the American Association of Variable Star Observers for providing help with the APASS dataset. This publication makes use of data products from the Two Micron All Sky Survey, which is a joint project of the University of Massachusetts and the Infrared Processing and Analysis Center/California Institute of Technology, funded by the National Aeronautics and Space Administration and the National Science Foundation. We thank Bernie Shiao, Rick White, and Brian McLean from the Barbara A. Mikulski Archive for Space Telescopes (MAST) at the Space Telescope Science Institute for their help in obtaining the Pan-STARRS1 catalogues. The Pan-STARRS1 Surveys (PS1) and the PS1 public science archive have been made possible through contributions by the Institute for Astronomy, the University of Hawaii, the Pan-STARRS Project Office, the Max-Planck Society and its participating institutes, the Max Planck Institute for Astronomy, Heidelberg and the Max Planck Institute for Extraterrestrial Physics, Garching, The Johns Hopkins University, Durham University, the University of Edinburgh, the Queen's University Belfast, the Harvard-Smithsonian Center for Astrophysics, the Las Cumbres Observatory Global Telescope Network Incorporated, the National Central University of Taiwan, the Space Telescope Science Institute, the National Aeronautics and Space Administration under Grant No. NNX08AR22G issued through the Planetary Science Division of the NASA Science Mission Directorate, the National Science Foundation Grant No. AST-1238877, the University of Maryland, Eotvos Lorand University (ELTE), the Los Alamos National Laboratory, and the Gordon and Betty Moore Foundation. We thank Michal Ko\v{c}er, then of the Kle\v{t} Observatory in the Czech Republic, for sharing copies of the UCAC4 astrometry index files. We thank the IT and technical staff at RSAA Mt Stromlo, in particular Gabe Bloxham, Joshua Rich, Bill Roberts, Kim Sebo, Annino Vaccarella and Col Vest, as well as Ian Adams and Peter Verwayen at Siding Spring Observatory, for persistent support on hardware, software and the telescope. We thank the staff at CDS, Strasbourg Observatory, France, for crucial support in creating the SkyMapper SkyViewer. This research made use of Astropy, a community-developed core Python package for Astronomy (Astropy Collaboration, 2013). We thank Roger Smith at Caltech and Mark Downing at ESO for suggestions related to the detector "tearing". We thank Lutz Wisotzki from AIP Potsdam for providing us with spectra from the Hamburg-ESO Survey. We also thank Helmut Jerjen, Anais M\"oller, and the anonymous referee for a very careful reading and comments improving the manuscript. We acknowledge the Gamilaroi people as the traditional owners of the land on which the SkyMapper Telescope stands.

\end{acknowledgements}

\begin{appendix}

\section{Source Extractor Parameters and TAP table schema}\label{appendix}

In Table~\ref{tab_sextractor}, we list the SExtractor parameters that are used and recorded in the DR1 photometry tables. 

In Table~\ref{tab_schemas}, we show the structure of tables in the various schemas of the SkyMapper ASVO TAP service. Each SkyMapper DR has its own schema, and external catalogues are collected into the 'ext' schema. Tables~\ref{tab_schema_master}-\ref{tab_schema_mosaic} provide descriptions of each column within the SkyMapper ASVO tables.

\onecolumn
\begin{longtable}{@{}ll@{}}
\caption{Source Extractor Parameters Measured}\\
\hline \hline
Parameter & Description\\
\hline
ALPHA\_SKY & Right Ascension (ICRS)\\
DELTA\_SKY & Declination (ICRS)\\
FLUX\_APER(10) & Fluxes in apertures$^a$\\
FLUXERR\_APER(10) & Flux errors in apertures$^a$\\
MAG\_APER(10) & Magnitudes in apertures$^a$\\
MAGERR\_APER(10) & Magnitude errors in apertures$^a$\\
FLUX\_AUTO & Flux in Kron aperture\\
FLUXERR\_AUTO & Flux error in Kron aperture\\
MAG\_AUTO & Magnitude in Kron aperture\\
MAGERR\_AUTO & Magnitude error in Kron aperture\\
KRON\_RADIUS & Kron radius in pixels\\
FLUX\_RADIUS(3) & Radii enclosing 20\%, 50\%, 90\% of total flux in Kron aperture\\
FLUX\_PETRO & Flux in Petrostian aperture\\
FLUXERR\_PETRO & Flux error in Petrosian aperture\\
MAG\_PETRO & Magnitude in Petrosian aperture\\
MAGERR\_PETRO & Magnitude error in Petrosian aperture\\
PETRO\_RADIUS & Petrosian radius in pixels\\
BACKGROUND & Background count level\\
FLUX\_MAX & Maximum per-pixel flux for the source above the background\\
MU\_MAX & Maximum surface brightness for the source\\
X\_IMAGE & $x$-axis barycentre in CCD frame\\
Y\_IMAGE & $y$-axis barycentre in CCD frame\\
A\_IMAGE & Semi-major axis length in pixels\\
ERRA\_IMAGE & Error in semi-major axis length\\
B\_IMAGE & Semi-minor axis length in pixels\\
ERRB\_IMAGE & Error in semi-minor axis length\\
THETA\_SKY & Position angle of semi-major axis (degrees East from North)\\
ELONGATION & Ratio of semi-major to semi-minor axis\\
FWHM\_IMAGE & Full width at half maximum in pixels, assuming a Gaussian profile\\
FLAGS & SExtractor flags for the source\\
CLASS\_STAR & Neural network-produced estimate of likelihood of point-like spatial profile\\
IMAFLAGS\_ISO & OR-combined flags from the image mask within the isophotal area for the source\\
NIMAFLAGS\_ISO & Number of pixels with flags from the image mask within the isophotal area for the source\\
\hline \hline
\label{tab_sextractor}
\footnotetext{$^a$The apertures used have diameters of 4, 6, 8, 10, 12, 16, 20, 30, 40, and 60 pixels.}
\end{longtable}
\twocolumn

\begin{table*}
\centering
\begin{threeparttable}[b]
\caption{Schemas and tables contained in the SkyMapper ASVO TAP service.}
\label{tab_schemas}
\begin{tabular}{lll}
\hline\hline       
Schema & Table & Description\\
\hline
dr1 & & Schema containing SkyMapper DR1 tables\tnote{a} \\
& master & Primary table of mean astrometric, photometric and shape measurements\\ & & per object, with cross-match information to external tables \\
& fs\_photometry & Table of individual photometric, astrometric and shape measurements\\ & & for every detection \\
& images & Table of exposures and telescope information \\
& ccds & Table of CCD-specific exposure information \\
& mosaic & Table of CCD positions in the mosaic \\
\hline
ext & & Schema containing external catalogues \\
& allwise & NASA Widefield Infrared Survey Explorer (WISE) AllWISE data release \\
& apass\_dr9 & AAVSO Photometric All Sky Survey (APASS) DR9 \\
& gaia\_dr1 & Gaia DR1 \\
& galex\_bcscat\_ais & BCS Catalog of Unique GALEX Sources - AIS \\
& galex\_bcscat\_mis & BCS Catalog of Unique GALEX Sources - MIS \\
& ps1\_dr1 & Pan-STARRS1 DR1 catalog, select columns for\\ & & decMean<=20 and nDetections>1 \\
& spec\_2dfgrs & 2 Degree Field Galaxy Redshift Survey (2dFGRS) \\
& spec\_2qz6qz & 2dF and 6dF QSO Redshift Surveys \\
& spec\_6dfgs  & 6 Degree Field Galaxy Survey (6dFGS) \\
& spec\_hesqso & Hamburg/ESO survey for bright QSOs \\
& twomass\_psc & Two Micron All Sky Survey (2MASS) Point Source Catalogue (PSC) \\
& twomass\_scn & Two Micron All Sky Survey (2MASS) Scan Table \\
& twomass\_xsc & Two Micron All Sky Survey (2MASS) Extended Source Catalogue (XSC) \\
& ucac4 & Fourth U.S. Naval Observatory CCD Astrograph Catalog (UCAC4) \\
& vsx & AAVSO International Variable Star Index \\
& yale\_bsc & Yale Bright Star Catalogue, 5th Revised Ed. \\
\hline
edr & & Schema containing SkyMapper Early Data Release tables \\
& master & Primary table of mean astrometric, photometric and shape measurements\\ & & per object, with cross-match information to external tables \\
& fs\_photometry & Table of individual photometric, astrometric and shape measurements\\ & & for every detection \\
& images & Table of exposures and telescope information \\
& ccds & Table of CCD-specific exposure information \\
\hline \hline
\end{tabular}
\begin{tablenotes}
\item[a] The table names listed here for the dr1 schema point to the latest versions of those tables (at present, DR1.1). Particular versions may be accessed by prepending, e.g., "dr1p0\_" (for DR1.0) or "dr1p1\_" (for DR1.1) to the table names within the dr1 schema.
\end{tablenotes}
\end{threeparttable}
\end{table*}

\onecolumn
\begin{longtable}{ll}
\caption{DR1 master Table Database Schema}\label{tab_schema_master}\\
\hline\hline
Column Name & Description\\
\hline
\endfirsthead
\caption* {DR1 master Table Database Schema (cont.)}\\
\hline\hline
Column Name & Description\\
\hline
\endhead
\endfoot
\hline\hline
\endlastfoot
smss\_j & SkyMapper Southern Survey designation of the form SMSS Jhhmmss.ss+/-ddmmss.s,\\
 & derived from mean ICRS coordinates\\
raj2000 & Mean ICRS Right Ascension of the object\\
dej2000 & Mean ICRS Declination of the object\\
e\_raj2000 & RMS variation around the mean Right Ascension, in milliarcseconds\\
e\_dej2000 & RMS variation around the mean Declination, in milliarcseconds\\
object\_id & Global unique object ID in the master table\\
mean\_epoch & Mean MJD epoch of the observations\\
rms\_epoch & RMS variation around the mean epoch\\
glon & Galactic longitude derived from ICRS coordinates. Not to be used as primary astrometric reference\\
glat & Galactic latitude derived from ICRS coordinates. Not to be used as primary astrometric reference\\
flags & Bitwise OR of Source Extractor flags across all observations\\
nimaflags & Total number of flagged pixels from bad, saturated, and crosstalk pixel masks across all observations\\
ngood & Number of observations used across all filters\\
ngood\_min & Minimum number of observations used in any filter (excluding filters with 0 observations)\\
nch\_max & Maximum number of child sources combined into this global object\_id in any filter\\
u\_flags & Bitwise OR of Source Extractor flags from u-band measurements in photometry table\\
u\_nimaflags & Number of flagged pixels from bad, saturated, and crosstalk pixel masks from u-band\\
 & measurements in photometry table\\
u\_ngood & Number of u-band observations used\\
u\_nch & Number of u-band child sources combined into this object\_id\\
u\_nvisit & Number of u-band observations for which the mean RA/Dec position fell on a CCD\\
v\_flags & Bitwise OR of Source Extractor flags from v-band measurements in photometry table\\
v\_nimaflags & Number of flagged pixels from v-band measurements in photometry table\\
v\_ngood & Number of v-band observations used\\
v\_nch & Number of v-band child sources combined into this object\_id\\
v\_nvisit & Number of v-band observations for which the mean RA/Dec position fell on a CCD\\
g\_flags & Bitwise OR of Source Extractor flags from g-band measurements in photometry table\\
g\_nimaflags & Number of flagged pixels from g-band measurements in photometry table\\
g\_ngood & Number of g-band observations used\\
g\_nch & Number of g-band child sources combined into this object\_id\\
g\_nvisit & Number of g-band observations for which the mean RA/Dec position fell on a CCD\\
r\_flags & Bitwise OR of Source Extractor flags from r-band measurements in photometry table\\
r\_nimaflags & Number of flagged pixels from r-band measurements in photometry table\\
r\_ngood & Number of r-band observations used\\
r\_nch & Number of r-band child sources combined into this object\_id\\
r\_nvisit & Number of r-band observations for which the mean RA/Dec position fell on a CCD\\
i\_flags & Bitwise OR of Source Extractor flags from i-band measurements in photometry table\\
i\_nimaflags & Number of flagged pixels from i-band measurements in photometry table\\
i\_ngood & Number of i-band observations used\\
i\_nch & Number of i-band child sources combined into this object\_id\\
i\_nvisit & Number of i-band observations for which the mean RA/Dec position fell on a CCD\\
z\_flags & Bitwise OR of Source Extractor flags from z-band measurements in photometry table\\
z\_nimaflags & Number of flagged pixels from z-band measurements in photometry table\\
z\_ngood & Number of z-band observations used\\
z\_nch & Number of z-band child sources combined into this object\_id\\
z\_nvisit & Number of z-band observations for which the mean RA/Dec position fell on a CCD\\
class\_star & Maximum stellarity index from photometry table (between 0=no star and 1=star)\\
radius\_petro & Mean r-band Petrosian radius\\
a & Mean r-band semi-major axis length\\
e\_a & Error in mean r-band semi-major axis length\\
b & Mean r-band semi-minor axis length\\
e\_b & Error in mean r-band semi-minor axis length\\
pa & Mean r-band position angle (E of N, -90..+90 deg)\\
e\_pa & Error in mean r-band position angle\\
u\_psf & Mean u-band PSF magnitude\\
e\_u\_psf & Error in u-band PSF magnitude\\
u\_petro & Mean u-band Petrosian magnitude\\
e\_u\_petro & Error in u-band Petrosian magnitude\\
v\_psf & Mean v-band PSF magnitude\\
e\_v\_psf & Error in v-band PSF magnitude\\
v\_petro & Mean v-band Petrosian magnitude\\
e\_v\_petro & Error in v-band Petrosian magnitude\\
g\_psf & Mean g-band PSF magnitude\\
e\_g\_psf & Error in g-band PSF magnitude\\
g\_petro & Mean g-band Petrosian magnitude\\
e\_g\_petro & Error in g-band Petrosian magnitude\\
r\_psf & Mean r-band PSF magnitude\\
e\_r\_psf & Error in r-band PSF magnitude\\
r\_petro & Mean r-band Petrosian magnitude\\
e\_r\_petro & Error in r-band Petrosian magnitude\\
i\_psf & Mean i-band PSF magnitude\\
e\_i\_psf & Error in i-band PSF magnitude\\
i\_petro & Mean i-band Petrosian magnitude\\
e\_i\_petro & Error in i-band Petrosian magnitude\\
z\_psf & Mean z-band PSF magnitude\\
e\_z\_psf & Error in z-band PSF magnitude\\
z\_petro & Mean z-band Petrosian magnitude\\
e\_z\_petro & Error in z-band Petrosian magnitude\\
ebmv\_sfd & E(B-V) from Schlegel+1998 extinction maps at the ICRS coordinates\\
prox & Distance to next-closest DR1 source\\
prox\_id & object\_id of next-closest DR1 source\\
edr\_id & object\_id of closest Early Data Release source\\
edr\_dist & Distance to closest Early Data Release source\\
twomass\_key1 & Unique identifier (pts\_key/ext\_key) of closest 2MASS source \\
twomass\_dist1 & Distance on sky to closest 2MASS source\\
twomass\_cat1 & 2MASS catalogue of closest match (point source PSC, or extended XSC)\\
twomass\_key2 & Unique identifier (pts\_key/ext\_key) of second closest 2MASS source \\
twomass\_dist2 & Distance on sky to second closest 2MASS source\\
twomass\_cat2 & 2MASS catalogue of second closest match (point source PSC, or extended XSC)\\
allwise\_cntr & Unique identifier (cntr) of closest AllWISE source \\
allwise\_dist & Distance on sky to closest AllWISE source\\
ucac4\_mpos & Unique identifier (mpos) of closest UCAC4 source \\
ucac4\_dist & Distance on sky to closest UCAC4 source\\
apass\_recno & Unique identifier (recno) of closest APASS DR9 source (only useful for this copy of APASS) \\
apass\_dist & Distance on sky to closest APASS DR9 source\\
gaia\_dr1\_id & Unique identifier (source\_id) of closest Gaia DR1 source\\
gaia\_dr1\_dist & Distance on sky to closest Gaia DR1 source\\
ps1\_dr1\_id & Unique identifier (objID) of closest Pan-STARRS1 DR1 source\\
ps1\_dr1\_dist & Distance on sky to closest Pan-STARRS1 DR1 source\\
galex\_bcs\_id & Unique identifier (objid) of closest GALEX BCSCat AIS source\\
galex\_bcs\_dist & Distance on sky to closest GALEX BCSCat AIS source\\
\hline
\end{longtable}
\twocolumn

\onecolumn
\begin{longtable}{ll}
\caption{DR1 fs\_photometry Table Database Schema}\label{tab_schema_fs_phot}\\
\hline\hline
Column Name & Description\\
\hline
\endfirsthead
\caption* {DR1 fs\_photometry Table Database Schema (cont.)}\\
\hline\hline
Column Name & Description\\
\hline
\endhead
\endfoot
\hline\hline
\endlastfoot
object\_id & Global object ID in master table associated with this observation.\\
object\_id\_local & Object ID in this (image,CCD) tuple. Not globally unique.\\
image\_id & Unique image ID as used by the Science Data Pipeline\\
ccd & CCD on which the object was detected (1-32)\\
filter & SkyMapper photometric filter (u, v, g, r, i, z)\\
ra\_img & ICRS Right Ascension of the object on this image\\
decl\_img & ICRS Declination of the object on this image\\
x\_img & CCD x-pixel coordinate of the object on this image\\
y\_img & CCD y-pixel coordinate of the object on this image\\
flags & Flags raised by Source Extractor\\
nimaflags & Number of flagged pixels from bad, saturated, and crosstalk pixel mask\\
imaflags & OR-combined flagged pixel types within isophotal radius from bad, saturated, and crosstalk pixel mask\\
background & Mean background level in counts\\
flux\_max & Peak count above background of the brightest object pixel\\
mu\_max & ZP-corrected peak surface brightness of object above background\\
class\_star & Source Extractor stellarity index (between 0=not star and 1=star)\\
a & Length of object semi-major axis in pixels\\
e\_a & Error in semi-major axis\\
b & Length of object semi-minor axis in pixels\\
e\_b & Error in semi-minor axis\\
pa & Object ellipse position angle on sky (E of N, -90..+90 deg)\\
e\_pa & Error in position angle\\
elong & Object elongation (a/b)\\
fwhm & Object full width at half maximum on detector\\
radius\_petro & Radius of Petrosian ellipse in units of a and b\\
radius\_kron & Radius of Kron ellipse in units of a and b\\
radius\_frac20 & Radius of circular aperture enclosing 20 per cent of total (Kron) flux\\
radius\_frac50 & Radius of circular aperture enclosing 50 per cent of total (Kron) flux (i.e. half-light radius)\\
radius\_frac90 & Radius of circular aperture enclosing 90 per cent of total (Kron) flux\\
chi2\_psf & Reduced Chi-squared of PSF fit\\
flux\_psf & Total counts under PSF fit\\
e\_flux\_psf & Error in PSF counts\\
mag\_psf & ZP-corrected PSF magnitude\\
e\_mag\_psf & Error in PSF magnitude\\
flux\_petro & Total counts within Petrosian ellipse\\
e\_flux\_petro & Error in Petrosian flux\\
mag\_petro & ZP-corrected Petrosian magnitude\\
e\_mag\_petro & Error in Petrosian magnitude\\
flux\_kron & Total counts within Kron ellipse\\
e\_flux\_kron & Error in Kron flux\\
mag\_kron & ZP-corrected Kron magnitude\\
e\_mag\_kron & Error in Kron magnitude\\
flux\_ap02 & Total counts within 2 arcsec diameter aperture centred on object\\
e\_flux\_ap02 & Error in 2 arcsec aperture flux\\
mag\_apc02 & ZP-corrected magnitude within 2 arcsec diameter aperture, corrected for PSF losses\\
e\_mag\_apc02 & Error in 2 arcsec aperture magnitude\\
flux\_ap03 & Total counts within 3 arcsec diameter aperture centred on object\\
e\_flux\_ap03 & Error in 3 arcsec aperture flux\\
mag\_apc03 & ZP-corrected magnitude within 3 arcsec diameter aperture, corrected for PSF losses\\
e\_mag\_apc03 & Error in 3 arcsec aperture magnitude\\
flux\_ap04 & Total counts within 4 arcsec diameter aperture centred on object\\
e\_flux\_ap04 & Error in 4 arcsec aperture flux\\
mag\_apc04 & ZP-corrected magnitude within 4 arcsec diameter aperture, corrected for PSF losses\\
e\_mag\_apc04 & Error in 4 arcsec aperture magnitude\\
flux\_ap05 & Total counts within 5 arcsec diameter aperture centred on object\\
e\_flux\_ap05 & Error in 5 arcsec aperture flux\\
mag\_apc05 & ZP-corrected magnitude within 5 arcsec diameter aperture, corrected for PSF losses\\
e\_mag\_apc05 & Error in 5 arcsec aperture magnitude\\
flux\_ap06 & Total counts within 6 arcsec diameter aperture centred on object\\
e\_flux\_ap06 & Error in 6 arcsec aperture flux\\
mag\_apc06 & ZP-corrected magnitude within 6 arcsec diameter aperture, corrected for PSF losses\\
e\_mag\_apc06 & Error in 6 arcsec aperture magnitude\\
flux\_ap08 & Total counts within 8 arcsec diameter aperture centred on object\\
e\_flux\_ap08 & Error in 8 arcsec aperture flux\\
mag\_apc08 & ZP-corrected magnitude within 8 arcsec diameter aperture, corrected for PSF losses\\
e\_mag\_apc08 & Error in 8 arcsec aperture magnitude\\
flux\_ap10 & Total counts within 10 arcsec diameter aperture centred on object\\
e\_flux\_ap10 & Error in 10 arcsec aperture flux\\
mag\_apc10 & ZP-corrected magnitude within 10 arcsec diameter aperture, corrected for PSF losses\\
e\_mag\_apc10 & Error in 10 arcsec aperture magnitude\\
flux\_ap15 & Total counts within 15 arcsec diameter aperture centred on object\\
e\_flux\_ap15 & Error in 15 arcsec aperture flux\\
mag\_apr15 & ZP-corrected magnitude within 15 arcsec diameter aperture, NOT corrected for PSF losses\\
e\_mag\_apr15 & Error in 15 arcsec aperture magnitude\\
flux\_ap20 & Total counts within 20 arcsec diameter aperture centred on object\\
e\_flux\_ap20 & Error in 20 arcsec aperture flux\\
mag\_apr20 & ZP-corrected magnitude within 20 arcsec diameter aperture, NOT corrected for PSF losses\\
e\_mag\_apr20 & Error in 20 arcsec aperture magnitude\\
flux\_ap30 & Total counts within 30 arcsec diameter aperture centred on object\\
e\_flux\_ap30 & Error in 30 arcsec aperture flux\\
mag\_apr30 & ZP-corrected magnitude within 30 arcsec diameter aperture, NOT corrected for PSF losses\\
e\_mag\_apr30 & Error in 30 arcsec aperture magnitude\\
\hline
\end{longtable}
\twocolumn

\onecolumn
\begin{table*}
\caption{DR1 ccds Table Database Schema} 
\centering
\begin{tabular*}{\textwidth}{@{}ll@{}}
\hline \hline
Column Name & Description\\
\hline
image\_id & Unique image ID as used by the Science Data Pipeline\\
ccd & CCD on which the object was detected (1-32)\\
filename & Filename of image\\
maskname & Filename of mask\\
image & Unique image\_id in text format\\
filter & Filter\\
mjd\_obs & MJD of the exposure\\
fwhm & Median full width at half maximum on the CCD\\
elong & Median elongation (A/B) on the CCD\\
nsatpix & Number of saturated pixels on the CCD\\
sb\_mag & Surface brightness estimate on the CCD\\
phot\_nstar & Number of objects found on the CCD\\
header & Image header\\
coverage & CCD RA/Dec coverage array\\
\hline \hline
\end{tabular*}\label{tab_schema_ccds}
\end{table*}

\begin{table*}
\caption{DR1 images Table Database Schema} 
\centering
\begin{tabular*}{\textwidth}{@{}ll@{}}
\hline \hline
Column Name & Description\\
\hline
image\_id & Unique image ID as used by the Science Data Pipeline\\
night\_mjd & Truncated Modified Julian Date at the start of night\\
date & Modified Julian Date at the start of the exposure\\
ra & Right Ascension of the exposure centre taken from the telescope\\
decl & Declination of the exposure centre taken from the telescope\\
field\_id & SkyMapper field ID\\
exp\_time & Exposure time\\
airmass & Airmass of the exposure centre taken from the telescope\\
filter & SkyMapper photometric filter (u, v, g, r, i, z)\\
rotator\_pos & Angular position of the instrument rotator\\
object & Survey/field/image\_type identifier (not unique)\\
image\_type & Type of image (fs=Fast/Shallow Survey, ms=Main Survey, std=Standard field)\\
fwhm & Average full width at half maximum for the exposure\\
elong & Average elongation (A/B) for the exposure\\
background & Average background counts for the exposure\\
zpapprox & Approximate photometric zeropoint for the exposure\\
\hline \hline
\end{tabular*}\label{tab_schema_images}
\end{table*}

\begin{table*}
\caption{DR1 mosaic Table Database Schema} 
\centering
\begin{tabular*}{\textwidth}{@{}ll@{}}
\hline \hline
Column Name & Description\\
\hline
ccd & CCD number (1-32)\\
x0 & Mosaic x-position of CCD pixel (0,0)\\
y0 & Mosaic y-position of CCD pixel (0,0)\\
\hline \hline
\end{tabular*}\label{tab_schema_mosaic}
\end{table*}
\twocolumn

\end{appendix}


\bibliographystyle{pasa-mnras}

\end{document}